\documentclass[letterpaper,11pt, nofootinbib]{article}
\pdfoutput=1
\usepackage{graphicx,psfrag,color}
\usepackage{pstricks,pst-node,pst-text,pst-3d}
\usepackage{float}
\usepackage{jheppub}
\usepackage{epsfig}
\usepackage{babel}
\usepackage{amsmath}
\usepackage{bbm}
\usepackage{graphicx}
\usepackage{appendix} 
\usepackage{capt-of}
\usepackage{tikz-cd}
\usepackage{comment}
\includecomment{myspecial}

\usepackage{subcaption}
\usepackage{wrapfig}
\usepackage{caption}
\usepackage[shortlabels]{enumitem}

\newcommand{\im}{\mathrm i}
\newcommand{\R}{\mathrm R}
\newcommand{\T}{\mathrm T}
\author[1]{Thiago Silva Tavares,}
\author[2]{Madhav Sinha,}
\author[3]{Linnea Grans-Samuelsson,}
\author[2]{Ananda Roy,}
\author[4,5]{and Hubert Saleur}
\affiliation[1]{Institute of Physics, University of S\~{a}o Paulo, S\~{a}o Paulo, SP, 05508-090, Brazil}
\affiliation[2]{Department of Physics and Astronomy, Rutgers University, Piscataway, NJ 08854-8019 USA}
\affiliation[3]{Microsoft Station Q, Santa Barbara, CA 93106-6105 USA}
\affiliation[4]{Institut de Physique Th\'eorique, Paris Saclay University, CEA, CNRS, F-91191 Gif-sur-Yvette}
\affiliation[5]{Physics Department, University of Southern California, Los Angeles, USA}

\emailAdd{tavares@df.ufscar.br, ms3066@physics.rutgers.edu,  linneag@microsoft.com, ananda.roy@physics.rutgers.edu, hubert.saleur@ipht.fr}

\abstract{Topological defect lines~(TDLs) in two-dimensional conformal field theories (CFTs) are standard examples of generalized symmetries in quantum field theory. Integrable lattice incarnations of these TDLs, such as those provided by spin/anyonic chains, provide a crucial playground to investigate their properties, both analytically and numerically. Here, a family of parameter-dependent integrable lattice models is presented, which realize different TDLs in a given CFT as the parameter is varied. These models are based on the general quantum-inverse scattering construction, and involve inhomogeneities of the spectral parameter. Both defect hamiltonians and (defect) line operators are obtained in closed form. By varying the inhomogeneities, renormalization group flows between different TDLs (such as the Verlinde lines associated  with the Virasoro primaries  $(1,s)$ and $(s,1)$ in diagonal minimal CFTs) are then studied  using  different aspects of the Bethe-ansatz as well as  ab-initio numerical techniques. Relationships with the anisotropic  Kondo model as well as its non-Hermitian version are briefly discussed. }

\title{Integrable RG Flows on Topological Defect Lines in 2D Conformal Field Theories}
\author{}
\date{July 2024}

\begin{document}

\maketitle
\section{Introduction}
\label{sec:intro}

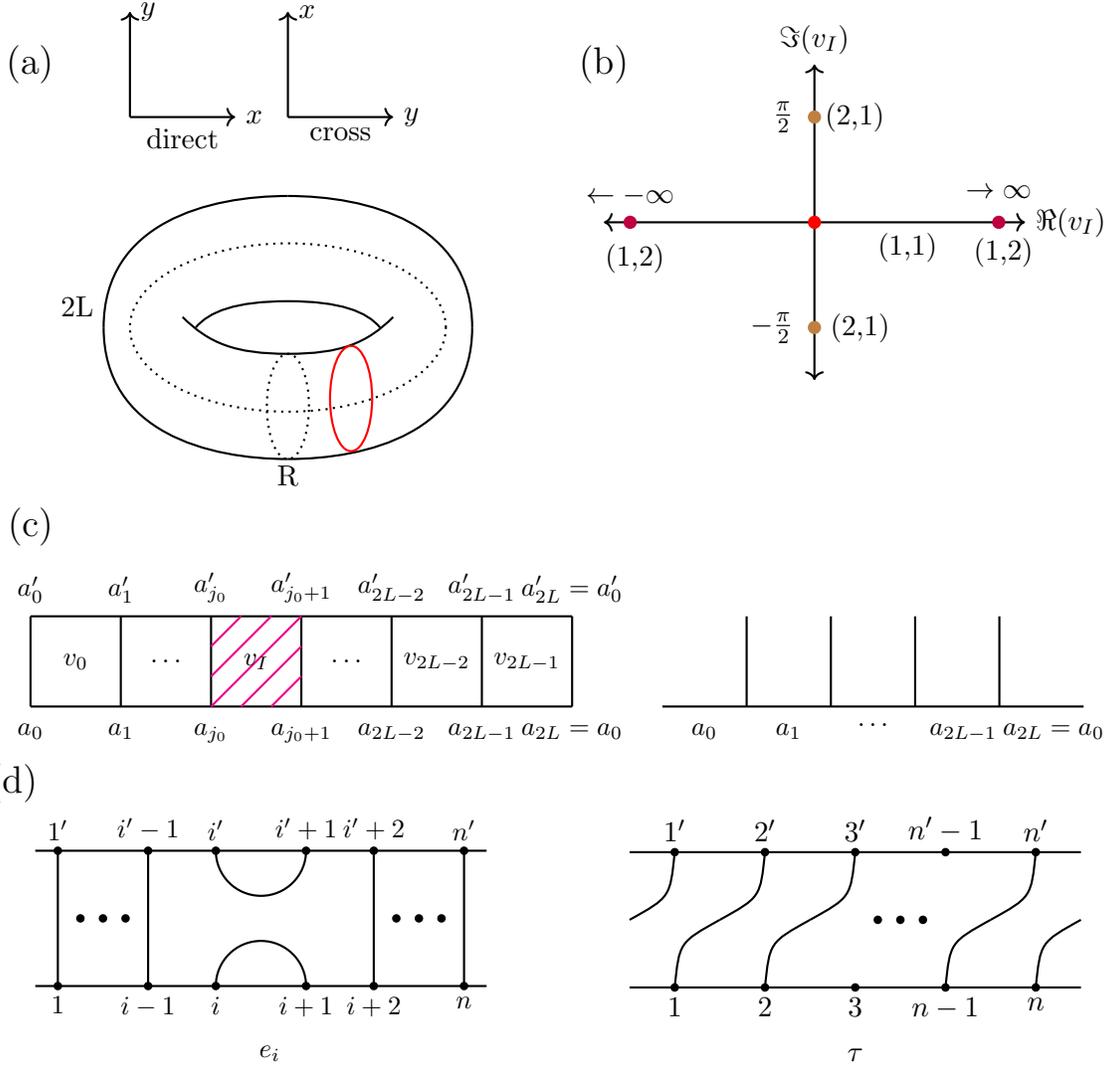
\begin{figure}[H]
\centering
\begin{tikzpicture}[thick, scale=0.7]
\node[anchor=center] at (9.1,3) {\Large{(a)}};

\begin{scope}[shift = {(14,-2)}]
\draw (-3.5,0) .. controls (-3.5,2) and (-1.5,2.5) .. (0,2.5);
\draw[xscale=-1] (-3.5,0) .. controls (-3.5,2) and (-1.5,2.5) .. (0,2.5);
\draw[rotate=180] (-3.5,0) .. controls (-3.5,2) and (-1.5,2.5) .. (0,2.5);
\draw[yscale=-1] (-3.5,0) .. controls (-3.5,2) and (-1.5,2.5) .. (0,2.5);

\draw (-2,.2) .. controls (-1.5,-0.3) and (-1,-0.5) .. (0,-.5) .. controls (1,-0.5) and (1.5,-0.3) .. (2,0.2);

\draw (-1.75,0) .. controls (-1.5,0.3) and (-1,0.5) .. (0,.5) .. controls (1,0.5) and (1.5,0.3) .. (1.75,0);
\draw[black, dotted] (0,-1.5) ellipse (0.4 cm and 1.0 cm);
\draw[red, thick] (1.2,-1.35) ellipse (0.4 cm and 1.0 cm);
\draw[black, dotted] (0.0,0) ellipse (3 cm and 1.6 cm);

\draw[<-] (0,6) -- (0,4) ;
\draw[->] (0,4) -- (2,4) ;

\draw[<-] (-3,6) -- (-3,4) ;
\draw[->] (-3,4) -- (-1,4) ;
\node[anchor=west] at (0,6) {$x$};
\node[anchor=west] at (2,4) {$y$};
\node[anchor=west] at (-3,6) {$y$};
\node[anchor=west] at (-1,4) {$x$};
\node[anchor = north] at (-2, 4) {direct} ; 
\node[anchor = north] at (1, 4) {cross} ; 
\node[anchor = south] at (0, -3.2) {R} ; 
\node[anchor = south] at (-4, 0) {2L} ; 
\end{scope}

\node[anchor=center] at (20,3) {\Large{(b)}};

    \draw [<->] (20,0) -- (28,0);
    \draw [<->] (24,3) -- (24,-3);
    \node[anchor=west] at (28,0) {$\Re(v_I)$};
    \node[anchor=south] at (24,3) {$\Im(v_I)$};

    \filldraw [purple] (20.5,0) circle (3pt)node[anchor=north] {};
    \node[anchor=north] at (20.5,-0.2){{ (1,2) }} ; 
    \node[anchor= north] at (27.5,0.9){ $\rightarrow \infty$ } ;
    \node[anchor= north] at (20.5,0.9){ $ \leftarrow  -\infty  $}  ;
    \node[anchor = east] at (23.8,2) {{$\frac{\pi}{2}$}} ; 
    \node[anchor = east] at (23.8,-2) {{$-\frac{\pi}{2}$}} ; 

    \filldraw [purple] (27.5,0) circle (3pt)node[anchor=north] {};
    \node[anchor=north] at (27.5,-0.1){{ (1,2) }} ; 
    \filldraw [brown] (24,2) circle (3pt)node[anchor=east] {};
    \node[anchor=west] at (24,2){ {(2,1) }} ; 
    \filldraw [brown] (24,-2) circle (3pt)node[anchor=east] {};
    \node[anchor=west] at (24.1,-2){{(2,1) }} ; 
    \filldraw [red] (24,0) circle (3pt)node[anchor=east] {};
    \node[anchor= north west] at (25,0){{(1,1)}} ; 

\end{tikzpicture}

            \begin{tikzpicture}[thick, scale=0.8, font = \small]
            \node[anchor=center] at (0,3) {\Large{(c)}};
            \draw[black, thick] (0,0) -- (9,0);
            \draw[black, thick] (0,1.5) -- (9,1.5);
            \draw[black, thick] (0,0) -- (0,1.5);
            \draw[black, thick] (1.5,0) -- (1.5,1.5);
            \draw[black, thick] (3,0) -- (3,1.5);
            \draw[black, thick] (4.5,0) -- (4.5,1.5);
            \draw[black, thick] (6,0) -- (6,1.5);
            \draw[black, thick] (7.5,0) -- (7.5,1.5);
            \draw[black, thick] (9,0) -- (9,1.5);
            \draw[magenta] (3,0) -- (4.5,1.5) ;
            \draw[magenta] (3.5,0) -- (4.5,1) ;
            \draw[magenta] (4,0) -- (4.5,0.5) ;
            \draw[magenta] (3,0.5) -- (4,1.5) ;
            \draw[magenta] (3,1) -- (3.5,1.5) ;
            \node[below] at (0,-0.1) {{$a_0$}};
            \node[below] at (1.5,-0.1) {{$a_1$}};
            \node[below] at (3,-0.1) {{$a_{j_0}$}};
            \node[below] at (4.5,-0.1) {{$a_{j_0+1}$}};
            \node[below] at (6,-0.1) {{$a_{2L-2}$}};
            \node[below] at (7.5,-0.1) {{$a_{2L-1}$}};
            \node[below] at (9,-0.1) {{$a_{2L} = a_0$}};
            \node[above] at (0,1.55) {{$a_0'$}};
            \node[above] at (1.5,1.55) {{$a_1'$}};
            \node[above] at (3,1.55) {{$a_{j_0}'$}};
            \node[above] at (4.5,1.55) {{$a_{j_0+1}'$}};
            \node[above] at (6,1.55) {{$a_{2L-2}'$}};
            \node[above] at (7.5,1.55) {{$a_{2L-1}'$}};
            \node[above] at (9,1.55) {{$a_{2L}' = a_0'$}};
            \node[] at (0.75,0.75) {{$v_0$}};
            \node[] at (2.25,0.75) {{$\ldots$}};
            \node[] at (3.75,0.75) {{$v_I$}};
            \node[] at (5.25,0.75) {{$\ldots$}};
            \node[] at (6.75,0.75) {{$v_{2L-2}$}};
            \node[] at (8.25,0.75) {{$v_{2L-1}$}};

            \draw[black, thick] (10.5,0) -- (17.5,0);
            \draw[black, thick] (11.9,0) -- (11.9,1.5) ;
            \draw[black, thick] (13.3,0) -- (13.3,1.5) ;
            \draw[black, thick] (14.7,0) -- (14.7,1.5) ;
            \draw[black, thick] (16.1,0) -- (16.1,1.5) ;
            
            \node[below ] at (11.2,-0.1) {{$a_0$}};
            \node[below  ] at (12.6,-0.1) {{$a_1$}};
            \node[below  ] at (14.0,-0.1) {{$\ldots$}};
            \node[below  ] at (15.5,-0.1) {$a_{2L-1}$};
            \node[below  ] at (17.0,-0.1) {{$a_{2L} = a_0$} };
           
    \end{tikzpicture}
\begin{subfigure}[b]{0.49\textwidth}
\begin{tikzpicture}[thick, scale=0.6, font = \small]
\node[anchor=center] at (-3.5,4.5) {\Large{(d)}};

\draw (-2.5,0) -- (-2.5,3);
\draw (-0.5,0) -- (-0.5,3);
\draw[black] (3,0) arc (0:180:1);
\draw[black] (3,3) arc (0:-180:1);

\draw (4.5,0) -- (4.5,3);


\draw (-3.,0) -- (7.,0);
\draw (-3.,3) -- (7.,3);
\filldraw [black] (-0.5,0) circle (2pt)node[anchor=north] {$i-1$};
\filldraw [black] (-0.5,3) circle (2pt)node[anchor=south] {{$i'-1$}};
\filldraw [black] (-2.5,0) circle (2pt)node[anchor=north] {$1$};
\filldraw [black] (-2.5,3) circle (2pt)node[anchor= south] {$1'$};
\filldraw [black] (1,0) circle (2pt)node[anchor=north] {$i$}; 
\filldraw [black] (1,3) circle (2pt)node[anchor=south] {$i'$}; 
\filldraw [black] (4.5,0) circle (2pt)node[anchor=north] {$i+2$}; 
\filldraw [black] (4.5,3) circle (2pt)node[anchor=south] {$i'+2$};
\filldraw [black] (3,0) circle (2pt)node[anchor=north] {$i+1$}; 
\filldraw [black] (3,3) circle (2pt)node[anchor=south] {$i'+1$};
\filldraw [black] (-2,1.5) circle (2pt);
\filldraw [black] (-1.5,1.5) circle (2pt);
\filldraw [black] (-1,1.5) circle (2pt);
\filldraw [black] (5,1.5) circle (2pt); 
\filldraw [black] (5.5,1.5) circle (2pt); 
\filldraw [black] (6,1.5) circle (2pt); 
 \draw (6.5,0) -- (6.5,3);
\filldraw [black] (6.5,3) circle (2pt)node[anchor=south] {$n'$};
\filldraw [black] (6.5,0) circle (2pt)node[anchor=north] {$n$};
\node[anchor = center] at (2.2,-1.5) {{$ {e_i}$}} ;
\end{tikzpicture}
\end{subfigure}
\begin{subfigure}[b]{0.49\textwidth}
    \begin{tikzpicture}[thick, scale=0.6]
\draw[white,thick] (-4,0) -- (-5.5,0);

\draw[black,thick] (-4,0) -- (6,0);
\draw[black,thick] (-4,3) -- (6,3);
\filldraw [black] (-3,0) circle (2pt)node[anchor=north] {$1$};
\filldraw [black] (-3,3) circle (2pt)node[anchor=south] {$1'$};
\filldraw [black] (-1,0) circle (2pt)node[anchor=north] {$2$};
\filldraw [black] (-1,3) circle (2pt)node[anchor=south] {$2'$};
\filldraw [black] (1,0) circle (2pt)node[anchor=north] {$3$};
\filldraw [black] (1,3) circle (2pt)node[anchor=south] {$3'$};
\filldraw [black] (3,0) circle (2pt)node[anchor=north] {$n-1$};
\filldraw [black] (3,3) circle (2pt)node[anchor=south] {$n'-1$};
\filldraw [black] (5,0) circle (2pt)node[anchor=north] {$n$};
\filldraw [black] (5,3) circle (2pt)node[anchor=south] {$n'$};

\draw (-3,0) .. controls (-2.9,1) .. (-2,1.5);
\draw (-2,1.5) .. controls (-1.1,2) .. (-1,3);

\draw (-1,0) .. controls (-0.9,1) .. (0,1.5);
\draw (0,1.5) .. controls (0.9,2) .. (1,3);

\filldraw [black] (1.5,1.5) circle (2pt); 
\filldraw [black] (2,1.5) circle (2pt); 
\filldraw [black] (2.5,1.5) circle (2pt); 

\draw (3,0) .. controls (3.1,1) .. (4,1.5);
\draw (4,1.5) .. controls (4.9,2) .. (5,3);

\draw (5,0) .. controls (5.1,1) .. (6,1.5);

\draw (-4,1.5) .. controls (-3.1,2) .. (-3,3);
\node[anchor = center] at (1,-1.5) {{$  {\tau}$}} ;

\end{tikzpicture}

\end{subfigure}

\caption{\label{fig_1} (a) Schematic of a TDL (in red)  for a two-dimensional CFT. In the direct channel, the TDL gives rise to a defect Hamiltonian $H_{{\cal D}}$. In the cross-channel, it acts instead as an operator ${\cal D}$ ($x$ is the space direction along the chain, while $y$ is the imaginary time direction). Note the same interpretation carries over to more general defect lines such as the perturbed TDLs~(b)  The different topological fixed points as a function of the real and imaginary parts of the inhomogeneity parameter. What happens away from the real and imaginary axis is discussed in section \ref{sec:fusion}.~ (c) Schematic of the faces of the resctricted solid-on-solid~(RSOS) models with inhomogeneity parameter~$v_I$ and the equivalent anyonic chain. ~(d) Graphical representation of the Temperley-Lieb generators~$e_i$ and  the lattice translation operator~$\tau$ acting on the state-space of the RSOS model.}

\end{figure}

Topological defect lines~(TDLs) in two-dimensional conformal field theories~(CFTs) maintain the continuity of the stress-energy tensor across the defect line~\cite{Petkova:2000ip, Fuchs:2002cm, Frohlich2004, Frohlich2006}. The associated line operator\footnote{In the recent literature, ``defect operator'' stands for the fields living at the end of TDLs, and not for $\widehat{\cal{D}}$,  which is instead usually refereed to as a line operator, or a defect line operator.}  ~$\widehat{\cal{D}}$
commutes with the Virasoro generators~[time direction perpendicular to the defect line in Fig.~\ref{fig_1}(a), referred to as ``cross- channel'']:
\begin{equation}
\label{eq:TDL_def}
\left[L_n, \widehat{\cal{D}}\right] = \left[\bar{L}_n, \widehat{\cal{D}}\right] = 0
\end{equation}
and as such, the TDLs can be deformed without affecting the values of the correlation functions as long as they are not taken across field insertions. 
Since these TDLs may or may not correspond to a global symmetry, they are not necessarily invertible, and thus, serve as paradigmatic examples of generalized symmetries in quantum field theories~\cite{Chang:2018iay, Cordova:2022ruw, Seiberg:2023cdc, Schafer-Nameki:2023jdn, Shao:2023gho, Apruzzi:2023uma, Bhardwaj:2023bbf, Seiberg:2024gek}. 

Identifying TDLs on the lattice is important for many reasons: it can unravel features of the  phase diagrams of models \cite{Frohlich2004,Feiguin:2006ydp} or open the route to constructing TDLs in non-rational CFTs \cite{jacobsen2023}. More importantly, it makes possible numerical investigation of subtle issues in particular regarding entanglement entropies \cite{Brehm2015,Brehm2016,Roy2021a}, allows further connection with aspects of category theory \cite{Aasen2016,Aasen:2020jwb}, and may make possible simulation of non-invertible symmetries on quantum devices \cite{Roy:2023wer}.

The simplest occurrence of TDLs in two-dimensional CFTs arises in the context of Virasoro minimal models - especially the diagonal ones, denoted here by ~${\cal M}(p+1, p )$, with central charge 
\begin{equation}
c=1-{6\over p(p+1)},~p=3,4\ldots
\end{equation}
known to describe the universality class of the $(p-1)^{\rm th}$ critical Ising model \cite{Roy:2023gpl}. It is well known \cite{Petkova:2000ip,petkova2000bcft} that in this case there are  as many  TDLs  (often called Verlinde lines in this case) as there are primary fields. The latter have conformal weights $\Delta=\bar{\Delta}=\Delta_{(rs)}$ with
\begin{equation}
\Delta_{(rs)}={\left[(p+1)r-ps\right]^2-1\over 4p(p+1)};1\leq r\leq p-1;1\leq s\leq p.
\end{equation}
In what follows, we  denote the corresponding TDL by ${\cal D}_{(rs)}$.\footnote{In some cases such as non-diagonal theories\cite{Sinha:2023hum}) one may distinguish chiral and antichiral defects: for diagonal RSOS models however the two behave identically \cite{belletete2020topological}.}

Lattice realizations - in particular integrable ones - are well known for all the ~${\cal M}(p + 1, p )$ CFTs. The TDLs ${\cal D}_{(1s)}$ have been build using different strategies in many references, first for small values of $p$ \cite{Grimm2001, Frohlich2004}, then for general $p$ \cite{Aasen2016,Belletete2023}. It is important to note that the construction  in these references give rise to defects that are topological {\sl on the lattice} already - meaning for instance that partition functions with lattice TDLs are invariant under moves of these lines compatible with the lattice geometry. In contrast, the identification of TDLs ${\cal D}_{(r1)}$ has been more difficult. No construction that would lead to topological invariance on the lattice is known at the present time: the proposal in \cite{belletete2020topological,Sinha:2023hum} (itself inspired from a construction in \cite{Chui_2003}) gives rise to TDLs only in the continuum limit. More precisely, the defect Hamiltonians  in \cite{belletete2020topological} are believed to describe a renormalization group (RG) flow from the $(1,s)$ defects to the $(s-1,1)$ ones, so the ${\cal D}_{(r1)}$ TDLs can only be obtained when the size of the system is not only much larger than the lattice spacing (which is the usual condition for obtaining the correct low-energy physics of the  CFT without defect) but also much larger than a crossover scale we shall represent as  $\T_I^{-1}$, $\T_I$ being a crossover temperature akin to the Kondo temperature in quantum impurity problems (see below for a discussion of the relationship with the  Kondo model). Note that once defects $(r,1)$ have been identified, all defects $(r,s)$ can also be obtained by  fusing the two types $(r,1)$ and $(1,s)$ on the lattice \cite{Sinha:2023hum,Sinha2024}. 

While in \cite{belletete2020topological,Sinha:2023hum,Sinha2024} the objective was to study the TDLs themselves,  one of the  goals of the present paper   is to quantitatively analyze the RG flows that make possible the realization of $(r,1)$ defects. The issue of RG flows on line defects is indeed interesting in itself. Known examples include Wilson or t'Hooft lines in four-dimensional gauge theories \cite{Kapustin_2006,Beccaria_2022}, or symmetry defects and impurities in three-dimensional quantum critical systems 
\cite{billó2013,Gaiotto_2014}. In two dimensions, line defects can be interpreted as  boundaries or interfaces, and also appear in the vast class of quantum-impurity problems. This latter point can be seen by using 
the  formulation [illustrated in  Fig.~\ref{fig_1}(a)] where (Euclidean)  time flows  parallel to the defect line - we shall refer to this situation as the ``direct channel''. In this case, impurity Hamiltonians associated with TDLs
provide nontrivial generalizations of multi-channel Kondo models~\cite{Ludwig1994, Affleck1995conformal, Fendley1999,Bachas_2004} with non-generic entanglement characteristics~\cite{Roy2021a, Rogerson:2022yim, Capizzi:2023vsz, Roy:2023wer}.

We consider the following  RG  flows linking two different TDLs. A bulk perturbation on a CFT with a TDL can trigger a flow to either a gapped or gapless phase. While the former is a topological quantum field theory~\cite{Baez1995, Schommer-Pries:2011rhj}, the latter is a different CFT with a different defect line~\cite{Chang:2018iay}. The characteristics of the IR theory are determined by the TDL in the UV CFT, manifesting a certain rigidity of RG flows between theories with generalized symmetries. A localized/defect perturbation, on the other hand, can trigger an RG flow between two different TDLs of the same CFT~\cite{Lesage:1998qf, Bachas_2010, Graham:2003nc, Kormos:2009sk}. These flows have been analyzed using perturbative and numerical techniques and constitute a more tractable special case of the more general RG flow between two conformal defect lines. The line operators for the latter obey the more general condition
\begin{equation}
\left[L_n - \bar{L}_n, \widehat{\cal{D}}\right] = 0.
\end{equation}
A full characterization of RG flows or the fixed points for the more general conformal defects remains an interesting open problem even for minimal models of two-dimensional CFTs. 

Here, a family of parameter-dependent restricted solid-on-solid~(RSOS) models is described that provide integrable lattice regularizations of the quantum field theory along the entire RG flow between two different TDLs in a given minimal model. While RSOS models have been used to realize TDLs in two-dimensional CFTs~\cite{Aasen2016, Belletete2023, Aasen:2020jwb, Belletete2023, Sinha:2023hum}, their utility in the characterization of the defect RG flows has remained unexplored. This work aims to fill this void. We calculate the change in the universal characteristics such as the defect g-function~\cite{Affleck1991} by computing thermodynamic characteristics as well as the variation of conformal dimensions along the flow. 
Note that while the focus of this work is on localized/defect perturbations, the presented lattice construction can be modified to consider bulk perturbations of TDLs following, for instance, Ref.~\cite{Reshetikhin:1993wm}.

The manuscript is organized as follows. In Sec.~\ref{sec: RSOS_def}, the scheme is presented for general RSOS models based on the Temperley-Lieb~(TL) algebra. Sec.~\ref{sec:TBA} provides details on the thermodynamic Bethe ansatz analysis of the flow connecting different TDLs. Sec.~\ref{sec:NLIE} describes the computation of the (defect) line  operator from the relevant transfer matrices. Secs.~\ref{sec:Ising} and~\ref{sec:pbig} provide the analysis of the lowest few  multicritical Ising models. Sec.~\ref{sec:fusion} discusses the topic of fusion of TDLs and the corresponding flows. Sec.~\ref{sec:concl} provides a concluding summary and outlook. 

\section{RSOS Models with a tunable inhomogeneity}
\label{sec: RSOS_def}
\subsection{Definitions}
We begin by collecting some well-known facts about RSOS models and their constructions using TL algebra (see, for example, Ref.~\cite{Saleur:1990uz}). For simplicity, we focus on the A-type models which lead to CFTs with diagonal modular invariants, but our construction can be straightforwardly generalized to non-diagonal models as well, see for example, Ref.~\cite{Sinha:2023hum} for results on the three-state Potts model.  

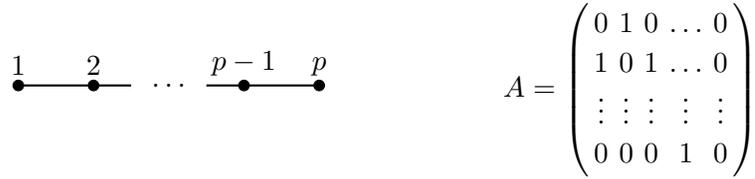
\begin{figure}[H]
\centering
\begin{subfigure}[b]{0.45\textwidth}
    \begin{tikzpicture}
    \draw[white] (0,0) -- (-3,0);
        \draw[black, thick] (0,0) -- (1.5,0) ;
        \draw[black, thick] (4,0) -- (2.5,0) ;
        \filldraw [black] (0,0) circle (2pt)node[anchor=south] {1};
        \filldraw [black] (1,0) circle (2pt)node[anchor=south] {$2$};
        \filldraw [black] (3,0) circle (2pt)node[anchor=south] {$p-1$};
        \filldraw [black] (4,0) circle (2pt)node[anchor=south] {$p$};
        \node[] at (2,0) {$\ldots$};
    \end{tikzpicture}
\end{subfigure}
\hspace{0.4cm}
\begin{subfigure}[b]{0.45\textwidth}
    \begin{equation*}
A = \begin{pmatrix}
        \, 0 & 1 & 0  & \ldots & 0  \, \\
        \, 1 & 0 & 1 & \ldots & 0   \,  \\
\, \vdots & \vdots & \vdots & \vdots & \vdots \, \\
        
        \, 0 & 0 & 0 & 1 & 0 \,  \\
\end{pmatrix}
\end{equation*}
\end{subfigure}
\caption{\label{fig_2} Schematic of the Dynkin diagram~$A_p$ and the corresponding adjacency matrix.}
\end{figure}

Recall that the heights of the RSOS model can be regarded as living on the nodes of the associated Dynkin diagram~[see Fig.~\ref{fig_2}] with the adjacency matrix~$A$. For the Dynkin diagram~$A_{p}$ with~$p$ nodes,~$p = 3, 4, \ldots$, the eigenvalues,~$\gamma^{(y)}$, and the associated eigenvectors,~$\phi^{y}_x$, are given as
\begin{equation}
\gamma^{(y)} = 2\cos\frac{y \pi}{p + 1}, \ \phi^{(y)}(x) = \sqrt{\frac{2}{p + 1}}\sin\frac{xy\pi}{p + 1},
\end{equation}
where~$x,y$ take integer values between $1, \ldots, p$. The TL generators are defined by the following relations:
\begin{align}
\label{eq:TL_gen}
e_j^2 = (q + q^{-1})e_j,\ e_ie_{i\pm1}e_i = e_i,\ [e_i,e_j] = 0,\ |i-j|\geq2,
\end{align}
with the parameter $q = e^{i\gamma}$ and~$\gamma \equiv \gamma^{(1)}= \pi/(p + 1)$. The matrix elements of the TL generators within two basis vectors of the RSOS Hilbert space is given as:
\begin{equation}
\langle x_1, \ldots, x_{2L}|e_i|x'_1,  \ldots, x'_{2L}\rangle = \prod_{j\neq i}\delta_{x_j, x'_j}\frac{\left[\phi^{(1)}(x_i)\phi^{(1)}(x'_i)\right]^{1/2}}{\phi^{(1)}(x_{i-1})}\delta_{x_{i-1}, x_{i+1}},
\end{equation}
where~$\phi^{(1)}$ stands for the eigenvector with the largest eigenvalue    and by $\phi^{(1)}(x_i)$ we mean the $x_i^{\rm th}$ entry of $\phi^{(1)}$.

Here, it is implicit that the state vectors~$|x_1, \ldots, x_{2L}\rangle$ are those which are allowed by the adjacency matrix of~$A_{n-1}$ Dynkin diagram. The TL generators can be used to construct the R-matrix for the six-vertex model~(see, for example, Ref.~\cite{Slavnov:2018kfx}):
\begin{equation}
\label{eq:R_def}
R_j(v_j)= \sin\left({\gamma\over 2} - i v_j\right) {\mathbb I} + \sin\left({\gamma\over 2}+i v_j\right)e_j,
\end{equation}
(note the $i$ factor in the arguments of the trigonometric functions) that satisfies the Yang-Baxter equation. It is convenient in what follows to use 
\begin{equation}
\label{eq:Rt_def}
\tilde{R}_j(v_j) = \sin\left({\gamma\over 2}+i v_j\right){\mathbb I} + \sin\left({\gamma\over 2} - i v_j\right) e_j , 
\end{equation}
 and the shift operator~$\tau$,
 which acts on the RSOS Hilbert space in the following way: 
\begin{equation}
\tau|x_1, \ldots, x_{2L}\rangle = |x_{2L}, x_1, \ldots, x_{2L - 1}\rangle.
\end{equation}

The lattice realization of TDLs (topological either on the lattice or only in the scaling limit) we study in this paper is obtained by considering a RSOS model homogeneous but for one (or maybe several when studying $r>2$ or $s>2$, see section \ref{sec:fusion}) column with a different (often referred to as ``impurity'') spectral parameter. See Figure \ref{fig_1}(c) for a schematic. Apart from this column, the spectral parameter is either chosen to be at the isotropic point $v=0\equiv v_{iso}$, or, for convenience and easier connection with other aspects of the problem, in the strongly isotropic Hamiltonian limit $v={i\gamma\over 2}$. 

In what follows,  we first  describe the model in the direct channel, presenting the relevant Hamiltonian with a localized impurity. This is followed by the discussion of the model in the cross-channel, where we focus on the transfer matrix that interpolates between different TDL operators. We focus on the case where the RG flow terminates at a single TDL in a given CFT defined on a spatial circle, but more complex examples could be considered using the same method. 

\subsection{Direct channel: RSOS Hamiltonian with a single defect}
\label{sec:RSOS_Ham}
The Hamiltonian for the RSOS model with a single defect located between sites $j_0$ and $j_0+1$ is given by~(see Ref.~\cite{Sinha2024} for derivation - in particular the Hamiltonians below are all normalized so that the ``sound velocity'' is equal to one):
\begin{align}
\label{eq:def_ham}
H(v_I) = -\frac{\gamma}{\pi\sin\gamma}\sum_{k=1}^{2L}e_k+\frac{\gamma}{\pi\sin\gamma}\frac{\sin (iv_I)}{\sin(\gamma + iv_I)}e_je_{j+1} -\frac{\gamma}{\pi\sin\gamma}\frac{\sin (iv_I)}{\sin(\gamma - iv_I)}e_{j+1}e_j, 
\end{align}
where~$v_I\in\mathbb{C}$ (the defect spectral parameter in the $R$ matrices is set to $v_j=\delta_{j,j_0} v_I + \im \gamma/2$)\footnote{The shift occurs due to our definition (\ref{eq:R_def}), which seems a bit cumbersome here but simplifies calculations in the cross-channel.}.

In our earlier  paper on the Potts model \cite{Sinha:2023hum} we considered three cases of particular interest.  
\begin{enumerate}[(i)]
\item $v_I = 0$: This is the case of the identity or the~$(11)$ defect. The resulting Hamiltonian is 
\begin{align}
\label{eq:H_I}
H|_{v_I = 0} = H_{\cal I}= -\frac{\gamma}{\pi\sin\gamma}\sum_{j = 1}^{2L}e_j,
\end{align}
which is the periodic RSOS Hamiltonian. In the scaling limit, the low-energy properties of this Hamiltonian are described by the minimal model~${\cal M}(p + 1, p )$~\cite{Andrews:1984af, Koo_1994}. 

\item $\Re (v_I)\rightarrow\pm\infty$: Then, the Hamiltonian realizes the Kramers-Wannier or the~$(12)$ defect line in~${\cal M}(p + 1, p)$ and is given by~\cite{Belletete2023, Sinha:2023hum}\footnote{Note that we have used the label $D$ for duality, instead of the label ${\cal D}$ for general defects.}:
\begin{align}
\label{eq:H_pmD}
H(v_I)|_{v_I\rightarrow\pm\infty} = H_{\pm D} = -\frac{\gamma}{\pi\sin\gamma}\sum_{j = 1}^{2L}e_j + \frac{\gamma}{\pi\sin\gamma}\left(q^{\pm1}e_{j_0}e_{j_0+1} + q^{\mp1}e_{j_0+1}e_{j_0}\right). 
\end{align}

Note that, for the A-type models considered in this work, $H_{\pm D}$ are isospectral. This can be seen by computing the partition function in the presence of the defect line~\cite{Petkova:2000ip}. Thus, it is sufficient to consider the case~$v_I\rightarrow\infty$. However, this is not the case in general; see, for example, the computation of the Kramers-Wannier defect line properties for the three-state Potts model in Ref.~\cite{Sinha:2023hum}. An important point to notice is that Hamiltonian (\ref{eq:H_pmD}) realizes the TDL already on the lattice \cite{Sinha2024}. 

\item $v_I = \pm i\pi/2$: The corresponding Hamiltonian is that for the~$(21)$ defect line in~${\cal M}(p+1, p )$ and is given as~\cite{Sinha:2023hum}:
\begin{equation}
\label{eq:H_vim}
H\left(v_I = \pm \frac{i\pi}{2}\right) = H_{D'} = -\frac{\gamma}{\pi\sin\gamma}\sum_{j = 1}^{2L}e_j + \frac{\gamma}{\pi\sin\gamma\cos\gamma}\left(e_{j_0}e_{j_0+1} + e_{j_0+1}e_{j_0}\right). 
\end{equation}
\end{enumerate}

As discussed in the case of the three-state Potts model in \cite{Sinha:2023hum}, this Hamiltonian does not correspond to a TDL on the lattice, but only in the scaling limit - that is at scales  much larger than the lattice spacing (so in particular $L\gg 1$).    

\bigskip

A given defect is characterized by the associated Affleck-Ludwig g-function~\cite{Affleck1991}. When viewed in the direct channel, the defect problem can be mapped into a boundary one by folding at the defect. The bulk CFT now has twice the central charge and all the well-known considerations of boundary CFTs can be subsequently applied.   
The g-factor for a defect of type $(rs)$ is  given by 
\begin{equation}
\label{eq:g_fac}
g_{(rs)}={S_{(rs)(11)}\over S_{(11)(11)}}= {\sin { \pi r\over p}\over 
\sin {\pi  \over p}}{\sin {\pi  s\over p+1}\over \sin{\pi \over p+1}},
\end{equation}
with $1\leq r\leq p-1,1\leq s\leq p$, and where $S$ is the modular S-matrix \cite{diFrancesco1997}. This leads to:
\begin{equation}
\label{eq:g_11_12_21}
g_{(11)} = 1, g_{(12)} = 2\cos\frac{\pi}{p + 1},\ g_{(21)} = 2\cos\frac{\pi}{p}.
\end{equation}
It is well known that in unitary theories, g-factors obey a g-theorem \cite{PhysRevLett.93.030402,Casini2016,Cuomo2022} and must decreases along an RG flow. It follows from this that 
the~$(11)$ TDL is the most stable, while the~$(12)$ is the most unstable. Perturbation of the (12) TDL in certain ways would lead to an RG flow to either the (11) or the (21) TDL. Around the~$v_I\rightarrow\infty$ point,~$H$ can be written as:
\begin{equation}
\label{eq:H_D_pert}
H(v_I) = H_{+D} +\frac{2i\gamma e^{-2v_I}}{\pi}\left(q^2e_{j_0}e_{j_0+1} - q^{-2}e_{j_0+1}e_{j_0}\right) + O(e^{-4v_I}).
\end{equation}
In the continuum limit, to leading order, this leads to a perturbed 
Hamiltonian of the form
\begin{equation}
\label{eq:H_pert}
H \approx H^*_{{\cal D}_{(12)}} + \Gamma\ \phi_{(13)}(0), 
\end{equation}
where~$H^*_{{\cal D}_{(12)}}$ is the CFT Hamiltonian for the (12) fixed point,~$\phi_{(13)}$ is the {\sl chiral} \footnote{Chiral perturbations are possible in this context since the perturbation is a defect field: such fields do not have to be mutually local.}  bulk primary field with conformal dimensions~$(\frac{p-1}{p+1},0)$ and~$\Gamma$ is the perturbing coupling. The perturbation is always relevant for finite~$p$ and turns marginal as~$p\rightarrow\infty$. The effects of this perturbation have been analyzed in Ref.~\cite{Kormos:2009sk} using perturbation theory and truncated conformal space approaches with the remarkable prediction that, depending on the sign\footnote{This is of course reminiscent of the effect of the sign of the coupling on bulk flows in minimal models \cite{Zamolodchikov1991b}.} of the coupling~$\Gamma$, the RG flow terminates at either the (11) or the (21) fixed point (it is  known that the conformal defect obtained as the IR fixed point in the case of a chiral perturbation such as (\ref{eq:H_pert}) is necessarily topological \cite{Kormos:2009sk}). Here, we provide a lattice incarnation of this flow by considering now the Hamiltonian (\ref{eq:def_ham}) away from the fixed points, and varying $v_I$ on two lines: $\Im(v_I)=0$ and $\Im(v_I)=\pm \pi/2$. (Note that the Hamiltonian (\ref{eq:def_ham}) is Hermitian on these two lines, but wouldn't be for arbitrary trajectories in the complex $v_I$ plane.) For large $\Re(v_I)$ we have the  identifications:
\begin{equation}
\Gamma \sim e^{-2v_I},\label{lambvrel}
\end{equation}
in the limit of large $v_I$. As~$\Re(v_I)$ varies from~$\infty$ to 0, the RG flow terminates at the~(11) or~(21) TDL depending on whether~$\Im(v_I) = 0$ or~$\Im(v_I) = \pm \pi/2$. In this respect, Hamiltonian  (\ref{eq:H_vim}) appears as a particular case of a manifold of defect Hamiltonians giving rise to the defect $(21)$ for large enough sizes.  Note that the lattice model is integrable for all choices of~$v_I$: this allows characterization of the entire RG flow using Bethe ansatz without resorting to perturbative techniques. 

Note that to properly observe the RG flow we need in fact two length scales. The coupling $\Gamma$ in (\ref{lambvrel}) provides the equivalent of the Kondo length scale $\R_I\propto \Gamma^{-{{p+1}\over 2}}$. The other length scale (call it $\R_{\rm Obs}$ for now) would depend on the property under study - for instance, the distance over which one would move the lattice defect line, or simply the size of the system - in particular, the size in the Euclidian time direction, which is the inverse of the temperature $\T={1\over \R}$. We expect results to depend on the ratio $\R_{\rm Obs}/\R_I$. When the ratio is large (and $\R_{\rm Obs},\R_I\gg1$), we expect to see properties of the IR fixed point of the flow, while when the ratio is small we should see properties of the UV fixed point instead. For $\R_{\rm Obs}/\R_I$ finite, we expect results in between the two fixed-points -  which is what we call exploring the RG flow.

The RG flow can be characterized by computing the defect g-functions as the parameter~$v_I$ is varied. In the direct channel, the g-function and its change along the flow can be obtained by computing the thermodynamic entropy at finite temperature. The change in the thermodynamic entropy yields the defect/impurity entropy, which equals the logarithm of the g-function along the flow. The change in the g-functions between different fixed points can also obtained by computing the entanglement entropy of a block located symmetrically around the defect and tracking the~${\cal O}(1)$ term~\cite{Calabrese2004, Calabrese2009}. For computations at the three aforementioned fixed points for the Ising and Potts cases, see Ref.~\cite{Roy2021a} and Ref.~\cite{Sinha:2023hum} respectively. However, the change in the entanglement entropy along the flow is, in general, different from the change in the thermodynamic entropy~\cite{Casini2016, Sela2017}, even though their values coincide at the fixed points. Here, we describe the change in the thermodynamic entropy, which can be directly compared to the Bethe ansatz results.

\subsection{Crossed channel: (defect) Line  Operator}
\label{sec:RSOS_Def_op}
While the above-described impurity Hamiltonians are interesting on their own and are amenable to exact analytical and numerical analysis, often, it is interesting to investigate the problem in the crossed channel. In this channel, a TDL acts as a line operator $\widehat{{\cal D}}$  on the Hilbert space of the periodic CFT, and commutes with the  two copies of the Virasoro algebra ~[Eq.~\eqref{eq:TDL_def}].  For a defect that is topological on the lattice, the corresponding defect operator commutes with the lattice discretizations of the  Virasoro generators \cite{Sinha:2023hum,Sinha2024}. In the case where the topological nature of the defect is recovered only in the scaling limit, commutations only occurs in this limit (see a discussion of this point for the three-state Potts model in \cite{Sinha:2023hum}). 

We now note that our formalism provides us with a natural line operator in the cross-channel - the latter is simply obtained within the  quantum inverse scattering method by inserting a row of RSOS faces with the (same) modified spectral parameter $v_I$. The corresponding line operator is then the transfer matrix $T(v_I)$ (also denoted $T^{(1)}(v_I)$ below), acting, since we are in the cross-channel, on a system of size $\R$ (which we also take even). Using simple manipulations it can be  written~\cite{Sinha2024}:
\begin{align}
\label{eq:T_def}
T(v_I) &= \sin (\gamma/2+iv_I)\tilde{R}_1(v_I)\tilde{R}_2(v_I)\ldots\tilde{R}_{\R-1}(v_I)\tau^{-1}\nonumber\\&\quad + \sin(\gamma/2 - iv_I)\tau R_{\R- 1}(v_I)R_{\R - 2}(v_I)\ldots R_1(v_I),
\end{align}
where~$R, \tilde{R}$ are defined in Eqs.~(\ref{eq:R_def}, \ref{eq:Rt_def}).
By construction, this object  belongs to the family of commuting transfer matrices, and  can be diagonalized simultaneously with the Hamiltonian of the periodic (since we are in the cross-channel) RSOS chain~[Eq.~\eqref{eq:H_I}]. Its eigenvalues can thus be computed using Bethe ansatz along the entire RG flow. 

In particular, the transfer matrix for the choices (\ref{eq:H_I}), (\ref{eq:H_pmD}) and (\ref{eq:H_vim})  leads to the corresponding defect operators:
\begin{align}
\widehat{\cal{D}}_{(11)} &\propto T(0),\ \ \widehat{\cal{D}}_{21} \propto T(\pm i\pi/2),\\
\widehat{\cal{D}}_{(12)} &\propto  \lim_{v_I\rightarrow\infty}\left[\frac{1}{\sinh v_I}\right]^{\R} T(v_I). \label{LatticeDefect12}
\end{align}
($\widehat{\cal{D}}_{(12)}$ is also called $Y$ in \cite{Belletete2023,Sinha:2023hum}).
Note that these expressions are  different from the ones used in \cite{Sinha:2023hum}.
It turns out indeed that at the fixed points,  different lattice  objects can realize the line operators in the continuum limit:  this will be discussed in more detail in \cite{Sinha2024}. 
Away from the fixed points, we will see below how  the~${\cal O}(1)$ term in the expectation value of~$T(v_I)$ provides the g-function along the RG flow. 
We will then compute it  using Bethe ansatz and density matrix renormalization group techniques for different models.  


Finally, we notice that, from the existence of the underlying family of commuting transfer matrices, $T(v_I)$ commutes with the Hamiltonian in the cross-channel. This agrees with the general expectation that the line operator ${\cal \widehat{D}}_{(12)}$
perturbed by the {\sl chiral} field $\phi_{(13)}$ still commutes with $L_0+\bar{L}_0$ in the perturbed CFT \cite{Runkel_2008,Kormos:2009sk}.

\subsection{Higher defects}

While we have mostly discussed the (12) defects and perturbations thereof, the construction can be rather easily generalized to (1s) defects. The TDLs ${\cal D}_{(1s)}$ can be obtained as topological objects on the lattice by using a generalization of the integrable approach used for ${\cal D}_{(12)}$ \cite{Sinha2024}. The basic idea is to introduce, instead of an elementary (spin $1$ in our notations) defect, a higher-spin $J$ defect. The latter is realized in the case of RSOS models by a modified row, where jumps of heights equal to $J,J-2,\ldots -J+2$ can occur, and the corresponding Boltzmann weights have an impurity spectral parameter \cite{Chui_2003, belletete2020topological,Sinha2024}. Equivalently - and somewhat more physically - one can introduce $j$ several (12) defects and use ``fusion'' to obtain a (1,1+J) one. There are different ways to implement this fusion (see \cite{Sinha2024}; see also below). The most elegant for our purpose involves  shifting the corresponding impurity spectral parameters. Let us illustrate this for the (13) defect. The strategy is to build an RSOS model where two neighboring rows have modified spectral parameters $v_{j_0},v_{j_0+1}$. Like before, we will focus on the Hamiltonian. Introducing the constant

\begin{equation}
C_j={\sin \gamma\over \sin(\gamma+iv_j)\sin(\gamma-iv_j)},
\end{equation}
the Hamiltonian with two defects reads

\begin{equation}\label{eq:two-imp-ham}
\begin{split}
    H^{(2)} = & -\frac{\gamma}{\pi \sin \gamma} \sum_{j \neq j_0, j_0 + 1, j_0 + 2 } e_j - \frac{\gamma}{\pi} C_{j_0} e_{j_0} - \frac{\gamma}{\pi} C_{j_0 + 1} R^{-1}_{j_0}\left(v_{j_0}+i{\gamma \over 2}\right) e_{j_0+1}R_{j_0}\left(v_{j_0}+i{\gamma \over 2}\right) 
    \\ 
   & - \frac{\gamma}{\pi \sin \gamma} R^{-1}_{j_0}\left(v_{j_0}+i{\gamma \over 2}\right)R^{-1}_{j_0 + 1}\left(v_{j_0+1}+i{\gamma \over 2}\right)e_{j_0 + 2}R_{j_0 + 1}\left(v_{j_0+1}+i{\gamma \over 2}\right)R_{j_0}\left(v_{j_0}+i{\gamma \over 2}\right).
    \end{split}
 \end{equation}
We now specialize to the choice \cite{KRS81}
\begin{eqnarray}
v_{j_0}=v_I+{i\gamma\over 2},\nonumber\\
v_{j_0+1}=v_I-{i\gamma\over 2}.\label{specparashift}
\end{eqnarray}
 Finally, introducing the Jones-Wenzl  projector $P_{j_0}^{(2)}=1-  \frac{e_{j_0}}{q+ q^{-1}}   $,  the final Hamiltonian of interest is 
\begin{equation}
H^{(J=2)}(v_I)=P_{j_0+1}^{(2)}H^{(2)}P_{j_0+1}^{(2)}=P_{j_0+1}^{(2)}H^{(2)}.
\end{equation}

Of course, this Hamiltonian could be written fully explicitly in terms of the Temperley-Lieb generators, but there is no need for this here. The main point is that in the cases $\Re(v_I)\to\pm\infty$ it reproduces $H_{{\cal D}_{(12)}}$ \cite{Sinha2024}, while with the choice $v_I=\pm i\pi/2$ it gives rise to $H_{{\cal D}_{(21)}}$. Around the $v_I\to\infty$ point, $H^{(J=2)}$ realizes, in the continuum limit, the perturbed defect 
\begin{equation}
H^{(J=2)}\approx H_{{\cal D}_{(13)}}+\Gamma \phi_{(13)}(0),
\end{equation}
generalizing (\ref{eq:H_pert})  for $J=1$. The case $\Im(v_I)=0$ flows to (12) while the case $\Im(v_I)=\pm \pi/2$ flows to (21). 

The same construction can be carried out in the cross-channel. Associated with a defect of spin $J$ we would need, in the integrable construction, the transfer (monodromy)  matrix describing the transport of a spin $J$ (the spin in the ``auxiliary space'' in the QISM language) through an array of spins $1$. The latter can be written directly, or obtained by the same fusion technique (see \cite{Belletete2023} for a discussion at the $(1,1+J)$ fixed point). We will study this object (denoted $T^{(J)}$) below. For completeness, we give here the corresponding matrix elements
\begin{equation}\label{fusedTmat}
\left(T^{(J)}(v_I)\right)_{\mathbf{x}}^{\mathbf{x}'}= \prod_{k=1}^{\R} ~^{(J)}W \! \! \left(\begin{smallmatrix} 
x'_{k} & x'_{k+1} \\  
x_{k} & x_{k+1} \end{smallmatrix} \Big| v_I  
\right),
\end{equation}
\begin{equation}
~^{(J)}W \! \! \left(\begin{smallmatrix} 
d & c \\  
a & b \end{smallmatrix} \Big| v \right) = 
(-1)^{(1+J+(b-d+c-a) (a-c)/4)}\sqrt{\frac{\phi^{(1)}(\frac{c+a-1-J}{2}) \phi^{(1)}(\frac{c+a+1+J}{2})}{\phi^{(1)}(b) \phi^{(1)}(d)}} \sin(\im v+ (b d-a c) \gamma/2), 
\end{equation}
for $c-b=d-a$,
\begin{equation}
~^{(J)}W \! \! \left(\begin{smallmatrix} 
d & c \\  
a & b \end{smallmatrix} \Big| v \right) =(-1)^{J ( (a+c-b-d)/4-1/2)}\sqrt{\frac{\phi^{(1)}(\frac{c-a+1+J}{2}) \phi^{(1)}(\frac{a-c+1+J}{2})}{\phi^{(1)}(b) \phi^{(1)}(d)}} \sin(\im v+ (a c-b d) \gamma/2),
\end{equation}
otherwise. The particular case $J=1$ of  corresponds to (\ref{eq:T_def}). Heights still take values of $A_p$, but now adjacency rules are a bit different. Since we are in the cross-channel, fusion is performed in the imaginary time direction, so heights $(a,b)$ and $(c,d)$ must still be neighbor on the diagram, while for heights $(a,d)$ and $(b,c)$ one must use the fused adjacency matrix $A^{(J)}$  obtained by fusion via $A^{(J+1)}= A A^{(J)}-A^{(J-1)}$, with $A^{(-1)}=0$. 

As a last comment, we also introduce the ``trivial'' transfer-matrix $T^{(0)}$ (useful in writing fusion relations, see below)
\begin{equation}
T^{(0)}(v_I)=\sinh^{\R}(v_I) ~ 1,
\end{equation}
so that all expressions here presented for $T^{(J)}(v_I)$ correspond to hermitian matrices for $\Im v_I=0,~\pi/2$. Moreover, the matrix elements are trigonometric polynomials of order $\R$ in the variable $i v_I$.

In the following we will denote the relevant eigenvalue of $T^{(J)}(v_I)$ by $\Lambda^{(J)}(v_I)$.

\section{Characterization of defect RG flows using Bethe Ansatz}
\subsection{Impurity Entropy using the Thermodynamic Bethe Ansatz}
\label{sec:TBA}

It is easy to write down the continuum limit thermodynamic Bethe ansatz (TBA) equations associated with our construction. The latter is based on an integrable array, which is a variant of the type discussed in \cite{Reshetikhin:1993wm} (for bulk scattering), and \cite{FENDLEY1994681} (for boundary scattering). The strategy consists in building the vacuum of the lattice model, then writing Bethe-ansatz equations for  the excitations over this vacuum, assuming the validity of the string hypothesis \footnote{While this validity seems established - as far as computing thermodynamic properties - in the case of the XXZ spin chain and RSOS models \cite{Hao_2013}, there are many cases where it is much less under control, and therefore, studying   alternative approaches seems certainly worthwhile, especially when some of the parameters are rational \cite{Gainutdinov_2016}.}  . Finally, one minimizes the free energy functional to obtain the TBA equations. 
The results hold for an infinite chain, and for temperatures and impurity coupling both going to zero, but in a finite ratio. In fact, these results are extremely close to those discussed in \cite{PhysRevLett.71.2485,Fendley_1996} in the context of the Kondo model, a fact we will comment on at the end of this paper. See also \cite{Lesage:1998qf} for a different approach in the context of roaming trajectories \cite{Zamolodchikov2006}. 

Let us now discuss the resulting ``physical'' \cite{Reshetikhin:1993wm} TBA.  For the lattice model defined on the $A_p$ Dynkin diagram, the diagram encoding this TBA is $A_{p-2}$ (so it consists of only one node for the Ising model), with incidence matrix $N_{mn}$. 

The impurity free-energy at temperature~$\T$ for a defect of (lattice) spin\footnote{We use conventions where the fundamental representation of $SU(2)$ has spin one.}  $J$ reads 

\begin{equation}
\text{f}_{I}^{(J)}=-\T\int {d\theta\over 2\pi} {\ln\left(1+e^{-\epsilon_{J}(\theta)}\right)\over \cosh(\theta-\theta_I)},\label{basicinti}
\end{equation}
where the $\epsilon$ are solutions of the TBA eqs
\begin{equation}
\epsilon_m(\theta)={1\over \T\cosh\theta}\delta_{m1}-\sum_n N_{mn} \int {d\theta'\over2\pi}{1\over \cosh(\theta-\theta')}\ln\left(1+e^{-\epsilon_n(\theta')}\right).\label{TBA}
\end{equation}
Variants of these equations are of course well known, and appear for instance in the TBA analysis of mininal models ${\cal M}(p+1,p)$ perturbed by the bulk $\Phi_{(13),(13)}$ operator \cite{Zamolodchikov1991a}. In  (\ref{basicint}), $\theta_I$ is an ``impurity rapidity'', which will turn out to be related with the inhomogeneity spectral parameter via
\begin{equation}
\theta_I=(p+1)v_I . 
\end{equation}
Note that equations (\ref{basicinti},\ref{TBA}) are written in the limit of an infinite system (chain). The scaling limit is obtained by taking $\T\to 0$ while the  rapidity $\theta\to\pm\infty$. These two limits correspond respectively to the left and right moving massless particles familiar in the massless scattering description of conformal field theories as UV fixed points of integrable flows \cite{Fendley:1993jh}. The inhomogeneity is then seen to interact only with left or right movers depending on whether $\theta_I$ itself tends to $\pm\infty$. Taking for instance the case $v_I\to\infty$ we have $\theta_I\to\infty$ and we are left, after absorbing $\T$ with a shift in the rapidity $\theta$,  with the TBA for right movers:
\begin{equation}
\epsilon_m(\theta)=2e^{-\theta}\delta_{m1}-\sum_n N_{mn} \int {d\theta'\over2\pi}{1\over \cosh(\theta-\theta')}\ln\left(1+e^{-\epsilon_n(\theta')}\right), \label{TBAeq}
\end{equation} 
together a slighlty modified version of (\ref{basicinti}):
\begin{equation}
\text{f}_{I}^{(J)}=-\T\int {d\theta\over 2\pi} {\ln\left(1+e^{-\epsilon_{J}(\theta)}\right)\over \cosh(\theta-\ln(\T/\T_I))},\label{basicint}
\end{equation}
where we have introduced the cross-over temperature $\T_I=e^{-\theta_I}$. Equations (\ref{TBAeq}, \ref{basicinti}) must   all
be considered in the limit 
 $\T\to 0,\theta\to\infty, \T/\T_I$ finite. The dimension of the perturbing operator follows from the periodicity  of the TBA system, and is the same as for the bulk perturbations considered in  \cite{Reshetikhin:1993wm}, i.e. $h=h_{13}={p-1\over p+1}$. It follows from scaling  that the relationship between $\T_I$ and the bare coupling $\Gamma$ should be:
 \begin{equation}
 \T_I\sim \Gamma^{1\over 1-h}\sim \Gamma^{p+1\over 2},
 \end{equation}
 so $\Gamma\sim e^{-2v_I}$, as found earlier in (\ref{lambvrel}). 

We analyze the  limiting behaviors of the impurity free-energy. Start with  $\theta_I=\infty$ so $\T_I=0$ and $\T/\T_I=\infty$ for all temperatures. This, we argued earlier, should  correspond to   the case of the $(1,1+J)$ defect. In this limit, only the region $\theta\to\infty$ will contribute to the integral (\ref{basicint}) , for which the source term of the Bethe equations will always be negligible. In this case therefore the TBA is temperature independent, and we can solve it immediately with
\begin{equation}
x_m=e^{-\epsilon_m}=\left({\sin\pi {(m+1)\over p+1}  \over \sin{\pi\over p+1}}\right)^2-1\label{firstx}.
\end{equation}
From this and (\ref{basicint}) we find  
\begin{equation}
\text{f}_{I}^{(J)}=-{\T\over 2}\ln (1+x_{J})=-\T\ln {\sin (1+J)\pi /(p+1)\over\sin\pi /p+1},~UV\label{resi}.
\end{equation}

We next consider the case $\theta_I$ finite. If $\T\to\infty$, nothing changes and we still get the result (\ref{resi}). In the other limit where  $\T\to 0$ on the other hand, since the integral (\ref{basicint}) is still dominated by the region $\theta$ near $\theta_I$, the  source term in (\ref{TBA}) will be very large, forcing $\epsilon_1=\infty$. This means the first node  disappears from the TBA. What remains a same system similar to the previous one,  but with one node less. We now have 
\begin{equation}
y_m=e^{-\epsilon_m}=\left({\sin\pi {m\over p}  \over \sin{\pi\over p}}\right)^2-1\label{secondx}.
\end{equation}
When going from (\ref{firstx}) to (\ref{secondx}) observe that both the numerator and the denominator have been shifted by one unit since the first node disappears (hence $p+1\to p$) but also the $m^{th}$ mode now becomes the $(m-1)^{th}$ one. We now obtain 
\begin{equation}
\text{f}_{I}^{(J)}=-{\T\over 2}\ln \left(1+y_{J}\right)=-\T\ln {\sin J\pi/p\over\sin\pi /p},~IR.
\end{equation}
Hence the two cases  we have identified in the limits $\T/\T_I\to \infty$ resp. $\T/\T_I\to 0$ correspond respectively to $g_{(1,1+J)}$ and $g_{(J,1)}$~[see Eq.~\eqref{eq:g_fac}]. In between these two values (that is, for $\T/\T_I$ finite) , the TBA should describe the RG flow in the scaling limit of the lattice model. 

We now observe that 
\begin{equation}
{g_{(1,1+J)}\over g_{(J,1)}}={g_{(1,1+\bar{J})}\over g_{(1+\bar{J},1)}},
\end{equation}
with 
\begin{equation}
\bar{J}\equiv p-1-J.
\end{equation}
This strongly suggests that, if we put the mass term on the last node instead of the first one, i.e. replace the system (\ref{TBA}) by the new system
\begin{equation}\label{TBAi}
\epsilon_m(\theta)= {2e^{-\theta}}\delta_{m,p-2}-\sum_n N_{mn} \int {d\theta'\over2\pi}{1\over \cosh(\theta-\theta')}\ln\left(1+e^{-\epsilon_n(\theta')}\right),
\end{equation}
while keeping the same formula (\ref{basicint}) for the impurity free-energy, we might describe the flow from $(1,1+\bar{J})$ to $(1+\bar{J},1)$. 

This claim can be put on firmer grounds in two ways. First, we can use the result mentioned in \cite{Reshetikhin:1993wm} that the RSOS $R$ matrices obey the following identity\footnote{The correspondence between conventions in the present paper and this reference is such that our $p+1$ is denoted $t$ there, while our spectral parameter $v$ is denoted  $u$ there.}
\begin{equation}
R^{JJ}\left(iv-{\pi\over 2}\right)=R^{J,\bar{J}}\left(iv \right).\label{Ridentity}
\end{equation}

Using that $\Gamma\propto e^{-2v_I}$ we see that the shift of ${\pi\over 2}$ in (\ref{Ridentity}) exactly corresponds to changing the sign of $\Gamma$. On the other hand, the TBA conjectured in (\ref{TBAi},\ref{basicint}) corresponds precisely to an impurity of spin $\bar{J}$ with real positive  coupling: we thus see that it maps exactly on the problem of an impurity of spin $J$ but with real negative coupling. Now it has been argued  in \cite{Graham:2003nc} that such coupling precisely induces a flow from $(1,1+\bar{J})$ to $(1+\bar{J},1)$, something we therefore confirm non-perturbatively here. The simplest case is $\bar{J}=1$, which corresponds therefore to a flow from $(1,2)$ to $(2,1)$. 

Another proof could be obtained by showing directly that the expansion of the free energy in powers of $\Gamma$ using the TBA (\ref{TBAi}) is the same as the expansion of the free energy using (\ref{TBA}) up to a change of sign of $\Gamma$. This could be done using the same arguments as those used by Zamolodchikov in \cite{Zamolodchikov1991b}. We will discuss this elsewhere. 

A final remark is that the TBA equations could of course be obtained by some educated guesswork for the impurity scattering directly in the integrable field theory limit, like e.g. what was done in \cite{Lesage:1998qf}. This too will be discussed elsewhere.

\subsection{Expectation value of line operators}
\label{sec:NLIE}
The derivation sketched in the previous paragraph and \cite{Reshetikhin:1993wm} is based on the string hypothesis and 
a variational procedure to obtain the free-energy. While fine in this very standard case, this strategy fails in other situations, especially in non-unitary cases such as the XXZ chain with defects (which we plan to discuss elsewhere). Luckily, there are other ways to proceed. One strategy would be to use the Quantum Transfer Method (QTM)~\cite{MSUZUKI1976,KLUMPER1993}, applied to the face models~\cite{TAVARES2023}, to express the statistical operator ${\rm e}^{-H/\T}$  in terms of a new object, called the {\sl quantum transfer matrix}. The latter  distinguishes itself from the usual row-to-row transfer matrix in that the Hamiltonian  is isolated from the other conserved charges through a suitable choice of the spectral parameters in the Trotter limit. This may not be sufficient especially (again) in non-unitary cases, where the other conserved charges beside the Hamiltonian may also play an important role. We thus prefer to stick with the row-to-row transfer matrix, and turn instead to the Non-Linear Integral Equations (NLIE) approach~\cite{KLUMPER1992}. This approach allows one to 
 extract the  defect entropy directly from finite-size studies. A little care must be taken however to do this properly, and obtain a universal result.

First, some elementary considerations. In the direct channel we calculate the free energy of a an infinite one-dimensional system with a defect, and obtain, since the free energy of the CFT is just proportional to the central charge times $\T^2$:
\begin{equation}
F=-\T\ln Z=-{c\pi\over 6} L{\T}^2+\text{f}_I(\T,\T_I),
\end{equation}
where $\text{f}_I$ is the impurity free-energy. We then deduce the impurity entropy from
\begin{equation}
s_I=-{\partial \text{f}_I\over \partial \T},
\end{equation}
We note that, if an energy term of $O(1)$ is added to the impurity Hamiltonian, it does not change the entropy. The latter is expected to be a universal quantity \cite{PhysRevLett.93.030402}, function of $\T/\T_I$  in the scaling limit: it is the quantity we are most interested in in this paper. 

In the crossed channel, the defect now acts like the  operator $T(v_I)$ (and coincides, up to a normalization,  with the TDL operator ${\cal D}$ at the fixed point), and we get the partition function, for large $L$
\begin{equation}
Z\sim \langle 0|T(v_I)|0\rangle \exp\left[{\pi c\over 6} L \T\right],
\end{equation}
where the exponential contribution arises from the usual form of the  Hamiltonian (with periodic boundary conditions) in this channel, $H^*=2\pi \T(L_0+\bar{L}_0-{c\over 12})$, and $|0\rangle$ is the corresponding ground-state.
Now an important point is that $\langle 0|T(v_I)|0\rangle$ does in general depend on $\T$ (of course it also depends on $\T_I$). Moreover, the normalization of the defect (line operator)  is always ambiguous: it could always be multiplied by a product of local interactions, leading to a renormalization of the partition function by an ``energy''  factor $\exp[-u(\T_I)/\T]$ (using that the ``length'' in this channel is $\R=1/\T$). 

Calculating the logarithm we get 
\begin{equation}
\text{f}_I(\T,\T_I)=-\T\ln \langle 0|T(v_I)|0\rangle,
\end{equation}
We now see that the entropy is in fact, setting $\omega\equiv \langle 0|T(v_I)|0\rangle$:
\begin{eqnarray}
s_I&=&{\partial\over \partial \T}\left(\T\ln \omega\right)\nonumber,\\
&=&\ln \omega-{1\over \T}{\partial\ln \omega\over \partial {1\over \T}}.
\label{correct}
\end{eqnarray}
instead of the naive $\ln\omega$. Among other things, we see that   if $\omega\to \exp[-u(\T_I)/\T]\omega$, $s_I\to s_I-{u\over \T}+{u\over \T}=s_I$ is indeed invariant under renormalizations of the line operator. While at the end points of the flow it is customary to think of the g-factor as $\langle 0|{\cal D}|0\rangle$ with a conformal normalization of ${\cal D}$ in mind \cite{Petkova:2000ip}, the universal quantity we will extract from the thermodynamics of the lattice model is not $\omega=\langle 0|T(v_I)|0\rangle$ but $s_I$, or what we will call the running g-factor defined as 
\begin{equation}
s_I\equiv \ln g(\T_I/\T).
\end{equation}
What we shall now do is determine the impurity free-energy (in fact, $\omega$)  using NLIE techniques, and then extract $s_I$ from (\ref{correct}).

The starting point of the NLIE approach is to write the fusion hierarchy \cite{BAZHANOV1989} of transfer matrices for the RSOS model (for a particularly lucid discussion see \cite{nepomechie2002}). This fusion hierarchy in turn becomes a set of functional relations for eigenvalues of these transfer matrices. Finally, these relations can be exploited using  analyticity properties \cite{KLUMPER1992}: for example, via  Fourier transform, one can re-express the different shifted functions in terms of  the same elementary analytical object,  leading to  a linear system of algebraic equations.  

It is convenient in our  case to eliminate the  eigenvalues in favor of the so-called Y-system functions
\begin{eqnarray}
y^{(m)}(v) &=& \frac{\Lambda^{(m-1)}(v) \Lambda^{(m+1)}(v)}{\Lambda^{(0)}(v-i\gamma\left({1+m \over 2}\right))\Lambda^{(0)}(v+i \gamma \left({1+m \over 2}\right))},  \nonumber
\\ 
Y^{(m)}(v) &=& \frac{\Lambda^{(m)}(v-i \frac{\gamma}{2}) \Lambda^{(m)}(v+i \frac{\gamma}{2})}{\Lambda^{(0)}(v-i \gamma \left({1+m \over 2}\right))\Lambda^{(0)}(v+i \gamma \left({1+m \over 2}\right))}, \label{ty-system}
\end{eqnarray}
(where $\Lambda^{(m)}$ are eigenvalues of the fused transfer matrix $T^{(m)}$ \cite{BAZHANOV1989,KLUMPER1992}) or directly use the relation $y^{(m)}(v-\im \gamma/2) y^{(m)}(v+\im \gamma/2)= Y^{(m-1)}(v) Y^{(m+1)}(v)$, $m=1,\ldots,p-2$. For the largest eigenvalue, the resulting NLIE  is\cite{KLUMPER1992}: 
\begin{equation}
\log y^{(m)}(v) = \R \log \left[\frac{{\rm e}^{(p+1) v} -1}{{\rm e}^{(p+1) v} +1} \right] \delta_{m1}+ \frac{1}{2 \pi} \sum_{m=1}^{p-2} N_{mn} \int \frac{p+1}{\cosh((p+1)(v-v'))} \log Y^{(n)}(v'){\rm d}v', \label{NLIE}
\end{equation}
where $Y^{(m)}(v)=1+y^{(m)}(v)$. One has then
\begin{equation}\label{Eigen01}
\log \Lambda^{({J})}(v_I) = \R  u^{({J })}(v_I) + \ln\omega^{(J)}(v_I), 
\end{equation}
where
\begin{equation}
u^{({J})}(v_I)=
-\int {\rm e}^{\im k v_I} \left[\frac{  \cosh \left( \frac{k \pi}{2(p+1)} \left(p-J \right)\right) -\cosh {\pi k \over 2 (p+1)}}{2 k \sinh\frac{k \pi}{2} \cosh \frac{k \pi }{2 (p+1)} }\right]{\rm d}k+\log \cosh(v_I), \label{bulk}
\end{equation}
 and
\begin{equation}\label{omega}
\ln\omega^{(J)}(v_I)= \frac{1}{2 \pi} \int \frac{p+1}{\cosh ((p+1)(v_I-v))} \log Y^{(J)}(v){\rm d}v. 
\end{equation}
where $Y^{(J)}$ can be obtained from the solution of (\ref{NLIE}). The term $u^{(J)}(v_I)$ is totally explicit, and consistent with  the normalization of the transfer matrix given in eq. (\ref{fusedTmat}).

Note that in this  approach, we focus on a system of size $\R$ in the ground-state, and obtain the continuum limit by taking the thermodynamic limit, i.e. letting  $\R \rightarrow \infty$. In particular, it is crucial to emphasize that the NLIE hold for any finite size $\R$. In the previous section we took a system already of infinite length $\L$, and did the thermodynamics to extract the  $\T \rightarrow 0$ behavior. Clearly, the NLIE approach is much more powerful for the question we are interested in.   

An important remark is in order regarding eq. (\ref{Eigen01}). Since $T(v_I)$ acts on a periodic system, one might be tempted to think that, as is familiar for transfer matrices of periodic physical models at their critical points, the logarithms of its eigenvalues should have no term of $O(1)$. This would indeed be the case if e.g. we considered the RSOS model at its isotropic point ({$v_{iso}=0$}) and evaluated $T(v_{iso})$ acting on its ground-state $|0\rangle$. Now the point is that, while the family of commuting transfer matrices has a common basis of eigenvectors, the ground-state vector may change as a function of the spectral parameter. What we do in the calculation (\ref{Eigen01}) is evaluate $T(v_I)$ not on its ground-state, but on the physical ground state, that is the ground-state of $T(v_{iso})$ (which happens to be the same as the ground-state of the Hamiltonian). There is then no physical reason that the corresponding eigenvalue should behave like the ground state of a physical conformal invariant system, and indeed it does not - and exhibits instead, typically, a term of $O(1)$ as in (\ref{Eigen01}).

While in \cite{KLUMPER1992,ZAMOLODCHIKOV1991} critical properties are the main concern, here we wish to obtain the exact flows and exponents when  perturbing the fixed point at $v_I=\infty$. Therefore, we re-scale the auxiliary functions by defining 
\begin{equation}\label{eq: auxil}
y_+^{(J)}(\theta)=y^{(J)}\left(\frac{\theta +\log \R}{p+1}\right),
\end{equation}
%
to obtain
\begin{equation}
\log y^{(m)}_+(\theta)= (-2 {\rm e}^{-\theta}+{\cal O}(\R^{-1})) \delta_{m1}+\frac{1}{2 \pi}\sum_{n=1}^{p-2} N_{mn} \int \frac{\log Y^{(n)}_+(\theta')}{\cosh(\theta -\theta')}{\rm d}\theta',
\end{equation}
with the $\omega$ factor given by
\begin{equation}
\omega^{(J)}(\theta_I)=\frac{1}{2 \pi} \int \frac{\log Y_+^{(J)}(\theta)}{\cosh (\theta+\log (\R/\R_I) )}{\rm d}\theta.
\end{equation}
Therefore, the large system-size limit for (in)finite $ \log \frac{\R_I}{\R} $ is well-defined, and these NLIE are formally equivalent to (\ref{TBAeq}) upon the identification $y_+^{(m)}(\theta)=\exp(-\epsilon^{(m)}(\theta))$. The  RG flow is obtained for finite values of  $\frac{\R_I}{\R}$,  interpolating between the fixed  points at $0$ and $\pm\infty$.  

Besides the fundamental eigenvalue $(J=1)$, we shall be interested in the flow of $J=p-2$. In fact, due to the reflection relation 
(\ref{Ridentity}), this $\omega$ function is nothing but $\omega^{(1)}(\theta_I+\pi \im(p+1)/2)$. Note that the latter function cannot be analytically continued to $\omega^{(1)}(\theta_I)$ with real $\theta_I$, because of the existence of branch cuts dividing the regions of analyticity in the $\R \rightarrow \infty$ limit. This is discussed more in section \ref{sec:fusion}.

A final remark: while $\R$ appears as a continuous parameter in the TBA and NLIE approaches, the comparison with lattice results only holds for $\R$ an even integer - which is the condition for  periodic RSOS model configurations to be properly defined. In a more general context - for instance for the study of defects in XXZ spin chains - more care would have to be taken in interepreting the NLIE eqs. for $\R$ continuous\footnote{This is, of course, related to Kramer's degeneracy in spin-chains, in that the NLIE are modified by the existence of the additional zeros of the largest eigenvalue inside the analytical strip, depending on the parity of the system size. However, in either case one can vary $\R$ continuously.}.


Having shown that the defect entropy and the $\omega$-factors can be obtained both from the TBA and the NLIE approach focussing on  the largest eigenvalues of the row-to-row transfer matrix,  we shall now  present results for these quantities along the RG flows, computed both for  finite systems and  in the scaling limit.

\section{$p = 3$: Ising CFT}
\label{sec:Ising}
In this section, we describe the impurity Hamiltonians and the corresponding line operators for the simplest case:~$p = 3$, {\it i.e.}, the Ising CFT. There are only three TDLs in this case: (11), (12) and (21). In the direct channel (see Fig.~\ref{fig_1}), the relevant Hamiltonians are given by those of the periodic, the duality-twisted Ising and the antiperiodic Ising chains respectively. Below, we show that our TL Hamiltonian~[Eq.~\eqref{eq:def_ham}] reduces to the expected results. A particularly nice feature of the Ising case is the fact that the impurity Hamiltonians for the three defect cases are free-fermionic in nature, which can be straightforwardly diagonalized. 
\subsection{Hamiltonians, partition functions and g functions}
The Ising representation for the TL algebra is:
\begin{equation}
\gamma = \frac{\pi}{4}, \ e_{2k-1} = \frac{1}{\sqrt{2}}\left(1 + \sigma_k^x\right), \ e_{2k} = \frac{1}{\sqrt{2}}\left(1 + \sigma_k^z\sigma_{k+1}^z\right). 
\end{equation}
Then, the expression for the parameter-dependent defect Hamiltonian reduces to:
\begin{align}
\label{eq:H_def_ising}
H(v_I) = H_{\cal I} +\frac{1}{2}\left[\frac{\sinh^2(v_I)}{\cosh(2v_I)}\left(1 + \sigma_{j_0}^x + \sigma_{j_0}^z\sigma_{j_0+1}^z\right) + \frac{\sinh(2v_I)}{2\cosh(2v_I)}\sigma_{j_0}^y\sigma_{j_0+1}^z\right],
\end{align}
where $H_{\cal I}$ is the periodic critical Ising model Hamiltonian:
\begin{equation}
\label{eq:Ising_pbc}
H_{\cal I} = -\frac{1}{4}\sum_{k = 1}^L\left(1 + \sigma_k^x\right) -\frac{1}{4}\sum_{k = 1}^L\left(1 + \sigma_k^z\sigma_{k+1}^z\right). 
\end{equation}
As expected,~$H(v_I=0) = H_{\cal I}$. In the limit~$\Re(v_I)\rightarrow\infty$, we recover the ``duality defect Hamiltonian''~\cite{Oshikawa1997, Grimm1992, Grimm2001}:
\begin{equation}
\label{eq:Ising_duality}
H_{\pm D} = H_{\cal I} + \frac{1}{4}\left(1 + \sigma_{j_0}^x + \sigma_{j_0}^z\sigma_{j_0+1}^z\pm\sigma_{j_0}^y\sigma_{j_0+1}^z\right), 
\end{equation}
where the $H_{\pm D}$ are related to each other by the local unitary operator~$\sigma_j^z$. Finally, for~$v_I = \pm i\pi/2$, the Hamiltonian reduces to:
\begin{align}
\label{eq:Ising_apbc}
H_{D'} &= H_{\cal I} + \frac{1}{2}\left(1 + \sigma_{j_0}^x + \sigma_{j_0}^z\sigma_{j_0+1}^z\right).
\end{align}
This Hamiltonian can be transformed to that of the antiperiodic Ising chain by the local unitary operator~$\sigma_j^z$. 

The partition functions for the Ising model with the three TDLs are well-known~\cite{Petkova:2000ip} and are useful for the verification of the spectrum obtained using finite-size computations. For the convenience of the reader, they are collected here:
\begin{align}
\label{eq:Ising_part_11}
Z_{11} &= \chi_{11}\chi_{11}^* + \chi_{12}\chi_{12}^* + \chi_{13}\chi_{13}^*,\\
\label{eq:Ising_part_12}
Z_{12} &= \chi_{12}\chi_{11}^* + \chi_{11}\chi_{12}^* + \chi_{13}\chi_{12}^* + \chi_{12}\chi_{13}^*,\\
\label{eq:Ising_part_21}
Z_{21} &= \chi_{13}\chi_{11}^* + \chi_{12}\chi_{12}^* + \chi_{11}\chi_{13}^*,
\end{align}
where the explicit form of the characters~$\chi_{r,s} = \chi_{r,s}(q_m)$ can be found, for instance, in Chap. 8 of Ref.~\cite{diFrancesco1997}. Here,~$q_m = e^{-2\pi \R/L}$ is the modular parameter~[see Fig.~\ref{fig_1}(a) for the definitions of $\R,L$]. The corresponding g-functions for the three Ising defects are given as:
\begin{equation}
g_{\left(11\right)} = g_{\left(21\right)} = 1,~~~~ g_{\left(12\right)} = \sqrt{2}.
\end{equation}
\subsection{Results: RG flow of scaling dimensions}
Next, we present the results for the energy spectrum of the impurity Hamiltonians discussed above. The spin-chain Hamiltonian of Eq.~\eqref{eq:H_def_ising} can be mapped to a fermionic one using the Jordan-Wigner transformation:
\begin{equation}
\gamma_{2k-1} = \left(\prod_{l<k}\sigma_l^z\right)\sigma_k^x, \ \gamma_{2k} = \left(\prod_{l<k}\sigma_l^z\right)\sigma_k^y,
\end{equation}
where~$\gamma_l$-s are real, fermionic operators satisfying:~$\{\gamma_k, \gamma_l\} = 2\delta_{k, l}$. Standard manipulations~\cite{Peschel2003, Latorre2004} lead to:
\begin{equation}\label{MajMaj}
H(v_I) = \frac{1}{2}H^{\text{Maj}} - L + \frac{1}{2}\frac{\sinh^2(v_I)}{\cosh(2v_I)},
\end{equation}
where 
\begin{align}\label{standardMaj}
H^{\text{Maj}} &= \frac{i}{2}\sum_{k = 1}^L\gamma_{2k-1}\gamma_{2k} + \frac{i}{2}\sum_{k = 1}^{L-1}\gamma_{2k}\gamma_{2k + 1}-\frac{iQ}{2}\gamma_{2L}\gamma_1\nonumber\\&\quad + \frac{\sinh^{2}(v_I)}{\cosh(2v_I)}\left(-i \gamma_{2j_0-1}\gamma_{2j_0}- i\gamma_{2j_0}\gamma_{2j_0+1}\right) + \frac{\sinh(2v_I)}{2\cosh(2v_I)}i\gamma_{2j_0-1}\gamma_{2j_0+1},
\end{align}
where~$Q = \prod_{k}\sigma_k^z$ is the conserved~$\mathbb{Z}_2$ charge. The Hamiltonian without defect is the usual
\begin{equation}
H^{\text{Maj}} = \frac{i}{2}\sum_{k = 1}^L\gamma_{2k-1}\gamma_{2k} + \frac{i}{2}\sum_{k = 1}^{L-1}\gamma_{2k}\gamma_{2k + 1}-\frac{iQ}{2}\gamma_{2L}\gamma_1,
\end{equation}
while the Hamiltonian with the (12) (duality) defect (obtained as $v_I\to\infty$) is  
\begin{align}\label{MajDual}
H^{\text{Maj}}_{{\cal D}_{(12)}} &= \frac{i}{2}\sum_{k = 1,k\neq j}^L\gamma_{2k-1}\gamma_{2k} + \frac{i}{2}\sum_{k = 1,k\neq j}^{L-1}\gamma_{2k}\gamma_{2k + 1}-\frac{iQ}{2}\gamma_{2L}\gamma_1,
\end{align}
so in the end it has the remarkable property that the Majorana mode $\gamma_{2j_0}$ is decoupled from the system.

For the (11) and (21) defects, the ground state lies in the~$Q = 1$ sector, while for the (12) defect, the ground state is two-fold (Kramer) degenerate: the two lowest states in the~$Q = \pm1$ sector have identical energies (which means, in the language of spins, that the ground state energies of the sectors odd and even under $\mathbb{Z}_2$ are identical in this case).


Perturbing Hamiltonian (\ref{MajDual}) by taking $v_I$ large but finite gives 
\begin{align}\label{MajDuali}
H^{\text{Maj}}_{{\cal D}_{(12)}} &\approx \frac{i}{2}\sum_{k = 1,k\neq j}^L\gamma_{2k-1}\gamma_{2k} + \frac{i}{2}\sum_{k = 1,k\neq j}^{L-1}\gamma_{2k}\gamma_{2k + 1}-\frac{iQ}{2}\gamma_{2L}\gamma_1\nonumber\\&\quad +ie^{-2v_I}\left (\gamma_{2j_0-1}\gamma_{2j_0}+\gamma_{2j_0}\gamma_{2j_0+1}\right)+{\cal O}(e^{-4v_I}).
\end{align}
Choosing $\Im(v_I)=0$ or $\Im(v_I)=\pm \pi/2$ simply changes the sign of the coupling as in eq. (\ref{eq:H_pert}). Otherwise, the interaction weakly couples the Majorana mode $\gamma_{2j}$ to its neighbours (in a way reminiscent to what happens for the resonant level model in the case of complex fermions)\footnote{ 
One can also check that when  $v_I=i\pi/2$ in (\ref{standardMaj}), one gets  the (21) defect  (antiperiodic Ising chain) after a unitary transformation.}.

\begin{figure}
\centering
\includegraphics[width = \textwidth]{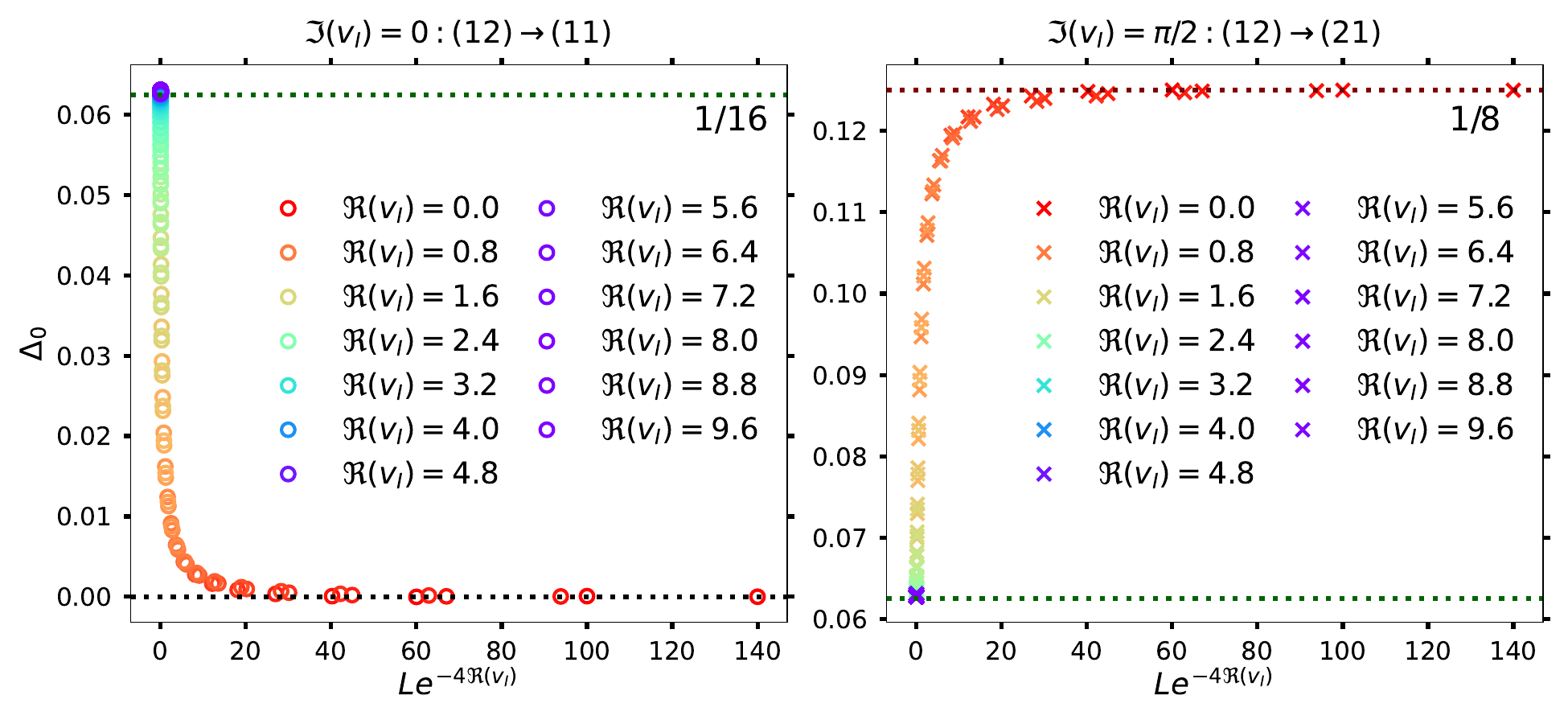}
\caption{\label{fig:TFI_GS} RG flow of the scaling dimension of the ground state for the Ising model. The defect Hamiltonian~[Eq.~\eqref{eq:H_def_ising}] was diagonalized using the free-fermion technique to obtain the ground state energy for~$L\in[40, 500]$ in steps of 10. The~$\Re(v_I)$ was chosen to vary between~0 and 10 in steps of 0.1. The left and right panels respectively show the results for the flow from the (12) to (11) and (12) to (21) TDLs.  The scaling dimension of the ground state~$\Delta_0$ is a function of~$L, v_I$. The data collapses to a single curve when plotted against the dimensionless variable:~$Le^{-4\Re(v_I)}$. The expected conformal dimensions at the (11), (12) and (21) fixed points are 0, 1/16 and 1/8 respectively~[see Eqs.~(\ref{eq:Ising_part_11}, \ref{eq:Ising_part_12}, \ref{eq:Ising_part_21})]. }
\end{figure}

The above Hamiltonian can be diagonalized to yield ground state energies as a function of system-size for different choices of $v$. The universal~${\cal O}(1/L)$ part of the ground state energy is obtained by fitting to the following expression:
\begin{equation}
E = E_0 L + E_1 -\frac{\pi }{6L}\left(c - 12\Delta_0\right) \, .\label{powerL}
\end{equation}
The scaling dimension for the ground state~$\Delta_0$ is, in general, a function of both~$L$ and~$v_I$. When plotted against the dimensionless parameter~$Le^{-4\Re(v_I)}$~[recall the perturbing operator in the Ising case has dimension 1/2, see Eq.~\eqref{eq:H_pert}]
, the data collapses into a single curve. This is shown in Fig.~\ref{fig:TFI_GS} for the RG flows from (12) to (11) and (12) to (21). The two different flows are obtained by choosing~$\Im(v_I)$ to be 0 and~$\pi/2$ respectively. The expected values for~$\Delta_0$ at the (11), (12) and (21) fixed points are given by 0, 1/16 and 1/8~~[see Eqs.~(\ref{eq:Ising_part_11}, \ref{eq:Ising_part_12}, \ref{eq:Ising_part_21})]. The obtained data is in good agreement with the expected predictions. 

\subsection{Results: RG flow of the impurity entropy}
\begin{figure}[H]
\centering
\includegraphics[width = \textwidth]{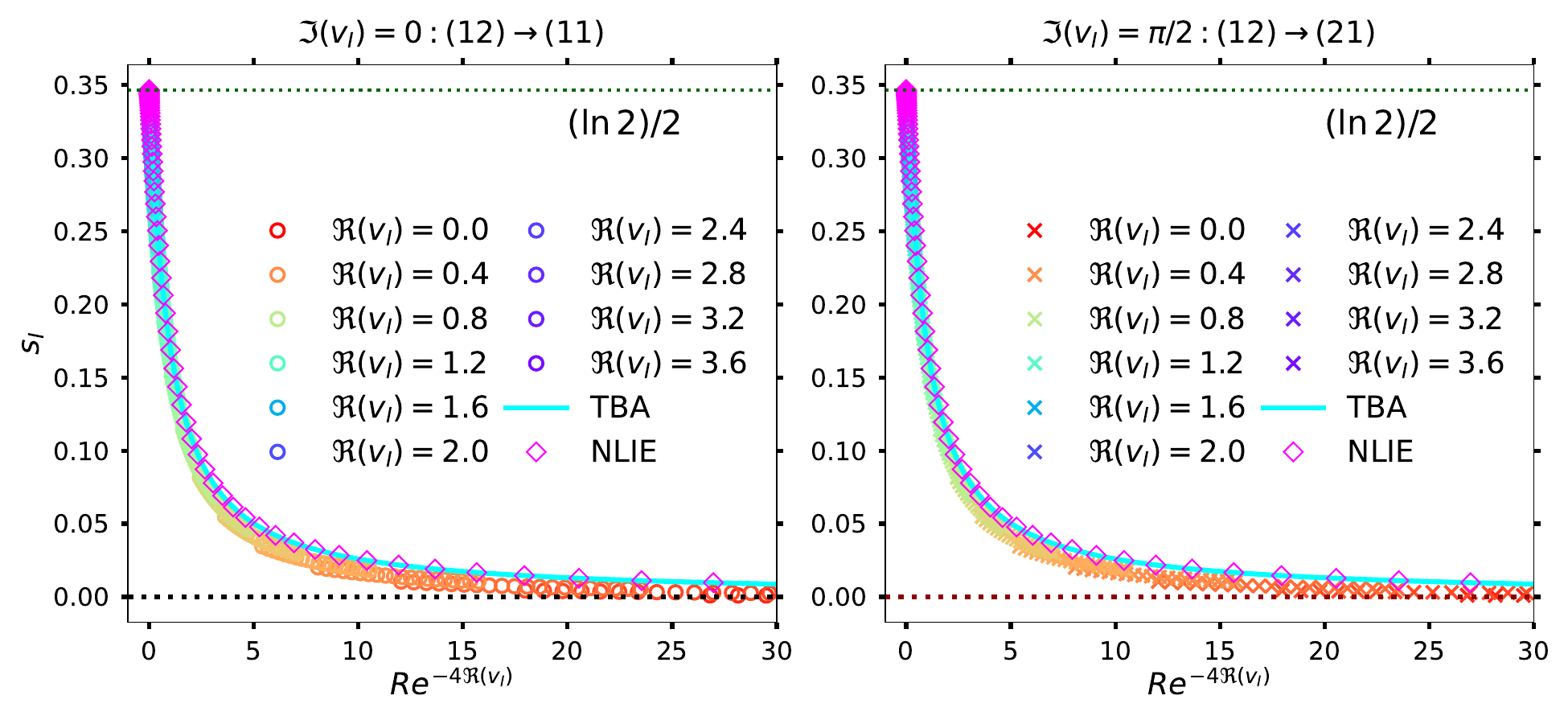}
\caption{\label{fig:TFI_TBA_FF} Change in the impurity entropy $s_I$ along the RG flow from (12) to (11), left panel, and from (12) to (21), right panel. The change was computed for~$\R = 1000$ with~$\Re(v_I)$ varying from 0 to 4 in steps of 0.1. In both cases, the g-function changes from~$\sqrt{2}$ to 1, leading to the shown change in the impurity entropy. For comparison, the analytical results computed using TBA technique are shown as the solid curve [Eq.~\eqref{eq:s_TBA_Ising}]. The change in the entropy can be computed further by computing the expectation value of the line operator using the row-to-row transfer matrix~(see Sec.~\ref{sec:NLIE}). The results of this computation is shown with purple diamond markers (label NLIE) on both panels. As seen from the two plots, the agreement between the ab-initio free-fermion, TBA and the transfer-matrix computations is reasonable.}
\end{figure}

The free fermionic nature of the Hamiltonians allow efficient computations of thermodynamic quantities such as the free energy and thermal entropy. If the eigenvalues of~$H_f$ are given by~$\lambda_k$,~$k = 1, \ldots, L$, up to irrelevant constants, the free energy,~$F$, is given by
\begin{equation}
\label{eq:F_Ising}
F(L, \R) = -\frac{1}{\R}\sum_{k = 1}^L\ln\left(2\cosh\frac{\R\lambda_k}{4}\right),
\end{equation}
where~$\R=1/\T$ is the inverse temperature. Thus, the thermal entropy,~$S$, is:
\begin{equation}
\label{eq:S_Ising}
S(\R, L) = \sum_{k=1}^L\ln\left(2\cosh\frac{\R\lambda_k}{4}\right) - \sum_{k = 1}^L\frac{\R\lambda_k}{4}\tanh\frac{\R\lambda_k}{4}.
\end{equation}
The change in impurity entropy along both of the aforementioned flows can be obtained by subtracting the entropy for~$\Re(v_I) = 0$ case. The corresponding numerical results obtained using the free-fermion method is shown in Fig.~\ref{fig:TFI_TBA_FF}. 

The numerical results are compared with the TBA results of Sec.~\ref{sec:TBA}. In this case, the TBA diagram has only one node, leading to the impurity free-energy
 in the scaling region:
\begin{align}
\text{f}_I &= \frac{1}{\R}\int \frac{d\theta}{2\pi}\frac{\ln\left(1 + e^{-2\R e^{-\theta}}\right)}{\cosh(\theta - \theta_I)},\nonumber\\
 &= \frac{1}{\pi \R}\int_0^\infty dx \frac{\ln\left(1 + e^{-2 \R x/ \R_I}\right)}{1+x^2}.
\end{align}
The defect entropy can be computed from the above and reads:
\begin{equation}
\label{eq:s_TBA_Ising}
s_I = \frac{1}{\pi}\int \frac{dx}{1+x^2}\left[\ln\left(1 + e^{-2\R x/ \R_I}\right)+\frac{2\R}{\R_I}\frac{x}{1+e^{2\R x/ \R_I}}\right].
\end{equation}
The change in the thermal entropy,~$\delta S$, shown with circles and crosses in Fig.~\ref{fig:TFI_TBA_FF}, should be compared to this impurity entropy~$s$, with the identification~$\R_I =  e^{\theta_I} = e^{4\Re(v_I)}$. The same change in the g-function can also be obtained by computing the expectation value of the transfer matrix~[Eq.~\eqref{eq:T_def}] in the ground state of the periodic Ising chain, see Sec.~\ref{sec:NLIE} for details. The results are shown with diamond markers in Fig.~\ref{fig:TFI_TBA_FF}. As seen from the two panels of the latter figure, the agreement between the ab-initio free-fermion, TBA and transfer-matrix computations is reasonable. 

The fact that  the TBA diagram has only one node leads us to expect that the flow of entropy from (12) to (11) and (12) to (21) are in fact numerically identical - something one can also check on Figure \ref{fig:TFI_TBA_FF}. This coincidence is expected from the form of the Hamiltonian in the Majorana language (\ref{MajDuali}). 

The degeneracy of the ground-state for the (12) defect arises from the existence of two Majorana zero-modes, one localized and one delocalized. As we flow towards (11) or (21), the localized zero-mode gets hybridized with the other modes, giving rise to a change of g-factor ${\sqrt{2}\over 2}\to 1$. As commented earlier, this can be considered a Majorana equivalent of the physics of the resonant level model. 

\section{ Results for $p\geq 4$} 
\label{sec:pbig}
\subsection{p = 4: Tricritical Ising model}
\label{sec:tci}
The computations done above for the Ising case can be performed for other minimal models. Here, we present the results for the tricritical Ising model, corresponding to~$p = 4$ in Eq.~\eqref{eq:def_ham}. 

Note that the Ising model is unique since it admits (due to coincidences in the Kac table) a realization of the $(21)$ defect that is topological on the lattice. For all other RSOS models (including the Tricritical Ising model), the $(21)$ defect is only topological in the scaling limit, and this is the limit we focus on. Below, we present the results for the scaling dimensions for the ground state in the direct channel and the expectation value of the line operator in the crossed channel. 

\begin{figure}
\centering
\includegraphics[width = \textwidth]{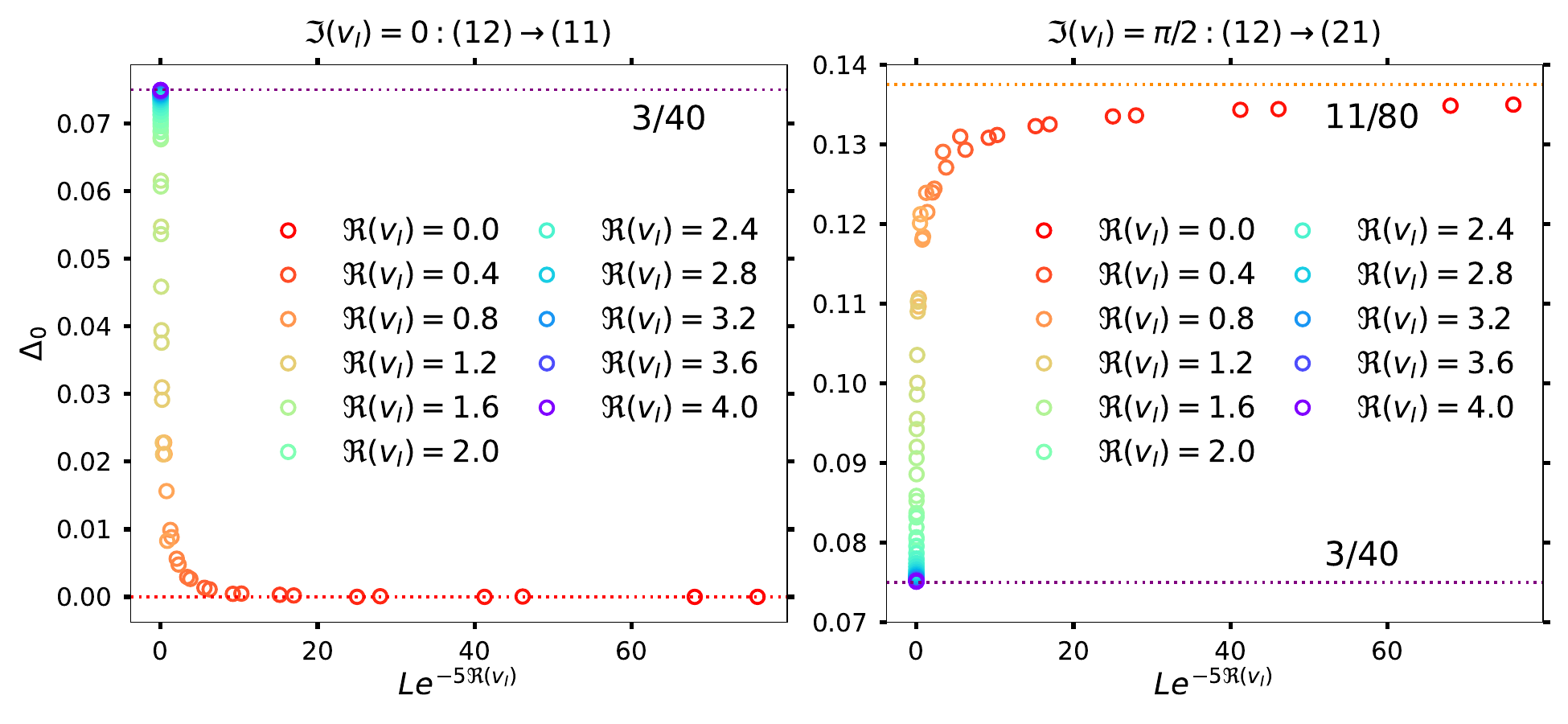}
\caption{\label{fig:TCI_GS} RG flows of the scaling dimension of the ground state for the tricritical Ising model with defect $J=1$. The ground state energy for the defect Hamiltonian~[Eq.~\eqref{eq:def_ham}] was obtained using the DMRG technique for~$L = 64$ to 80 in steps of 2.~$\Re(v_I)$ was chosen to vary between 0 and 4.0 in steps of 0.1. The left and right panels respectively show the results for the flow from the (12) to (11) and (12) to (21) TDLs   The scaling dimension of the ground state~$\Delta_0$ is a function of~$L, v_I$. The data collapses to a single curve when plotted against the dimensionless variable:~$Le^{-5\Re(v_I)}$. The expected conformal dimensions at the (11), (12) and (21) fixed points are 0, 3/40 and 11/80 respectively. }
\end{figure}

Fig.~\ref{fig:TCI_GS} shows the results for the scaling dimensions for the ground state energy. The partition functions are easily obtained using the results of Ref.~\cite{Petkova:2000ip} and are not shown here for brevity. The scaling dimensions for the ground state,~$\Delta_0$, in the presence of the (11), (12) and (21) defects are 0, 3/40 and 11/80 respectively. The flow of ~$\Delta_0$ between the three TDLs is obtained by computing the ground state energy of the parameter dependent Hamiltonian in Eq.~\eqref{eq:def_ham}. Since the~$p = 4$ model is not free-fermionic, the ground state energy is obtained using the density matrix renormalization group~(DMRG) technique. This is done for system-sizes~$L$ ranging from 60 to 80 in steps of 2 with~$\Re(v_I)$ varying between 0 and~4 in steps of 0.1. Note that the system-sizes we are able to analyze are much smaller  than for the Ising case since the DMRG computations are substantially more resource and time consuming. 
To improve the precision in the obtained conformal dimension, we used the fact that the extensive part of the energy~$E_0$~[Eq.\eqref{powerL}] is known exactly~\cite{Koo_1994}:
\begin{eqnarray}
    E_0 = \frac{\gamma}{\pi}\int_{-\infty}^{\infty} dt\frac{\sinh\left(\pi - \gamma\right)t}{\sinh(\pi t)\cosh(\gamma t)},
\end{eqnarray}
where~$\gamma$ is defined below Eq.~\eqref{eq:TL_gen}. Conformal dimensions were then obtained by fitting energies to the usual expectation from field theory, including phenomenological  higher order (in ${\cal O}(1/L)$) corrections:
\begin{equation}
    E = E_0L+E_1 -\frac{\pi}{12L}(c - 12\Delta_0) + E_3/L^2 + E_4/L^3.
\end{equation}
Notice that the finite-size scaling involves a length $2L$ (in lattice units)  for our system, compatible with the normalization of the  Hamiltonian (\ref{eq:H_I}). Formulas for the Ising model (\ref{powerL}) involved $L$ instead because, when going to the Majorana formulation, an extra factor ${1\over2}$ was introduced in eq. (\ref{MajMaj}) to match with the existing literature. 
As shown in Fig.~\ref{fig:TCI_GS}, the agreement between the obtained conformal dimensions at the fixed points and the expected results is reasonable. We are unaware of an analytical prediction for the entire flow of the conformal dimensions and we leave this interesting problem for the future. 

\begin{figure}[H]
\centering
\includegraphics[width = \textwidth]{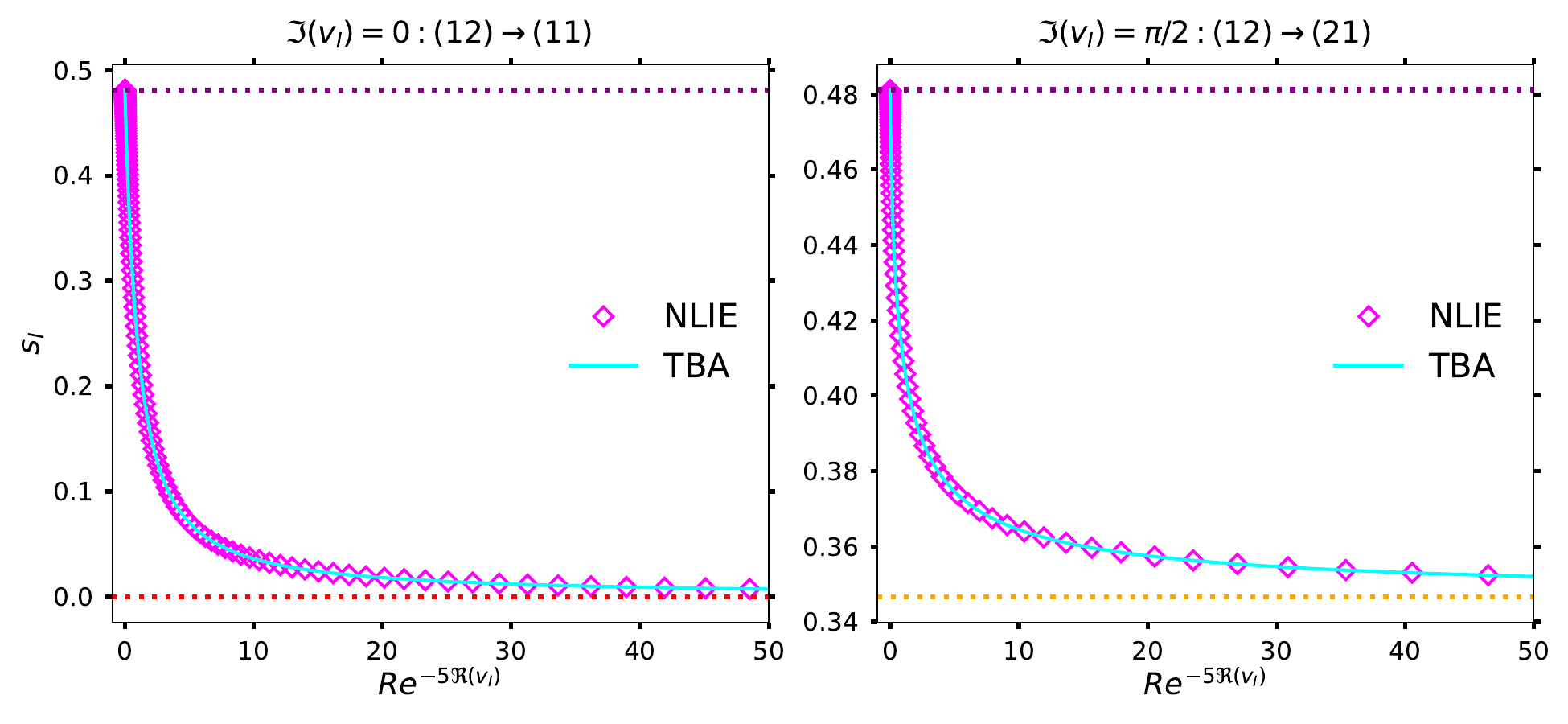}
\caption{\label{fig:TBA_A4_flow} Change in the impurity entropy for the tricritical Ising model  along the RG flow from (12) to (11), left panel, and from (12) to (21), right panel. The impurity entropies vary between the expected values~[Eq.~\eqref{eq:g_11_12_21}] with the TBA results~(solid line) agreeing reasonably well with the  transfer-matrix ones~(diamond markers, labelled NLIE). For the latter we have used $\R =1000$. }
\end{figure}

As in the Ising case, the change in the impurity entropy between the different fixed points can be obtained by computing the expectation value of the line operator as well as TBA. The results are shown in Fig.~\ref{fig:TBA_A4_flow}. The impurity entropy varies between the values of~$\ln[2\cos(\pi/5)]$ and 0 for the flow between the (12) and (11) TDLs. The corresponding variation for the flow between (12) and (21) TDLs is between~$\ln[2\cos(\pi/5)]$ and~$\ln[2\cos(\pi/4)]$~[see Eq.~\eqref{eq:g_11_12_21}]. 

\begin{figure}[H]
\begin{minipage}{0.5\linewidth}
\begin{center}
\includegraphics[width= \linewidth]{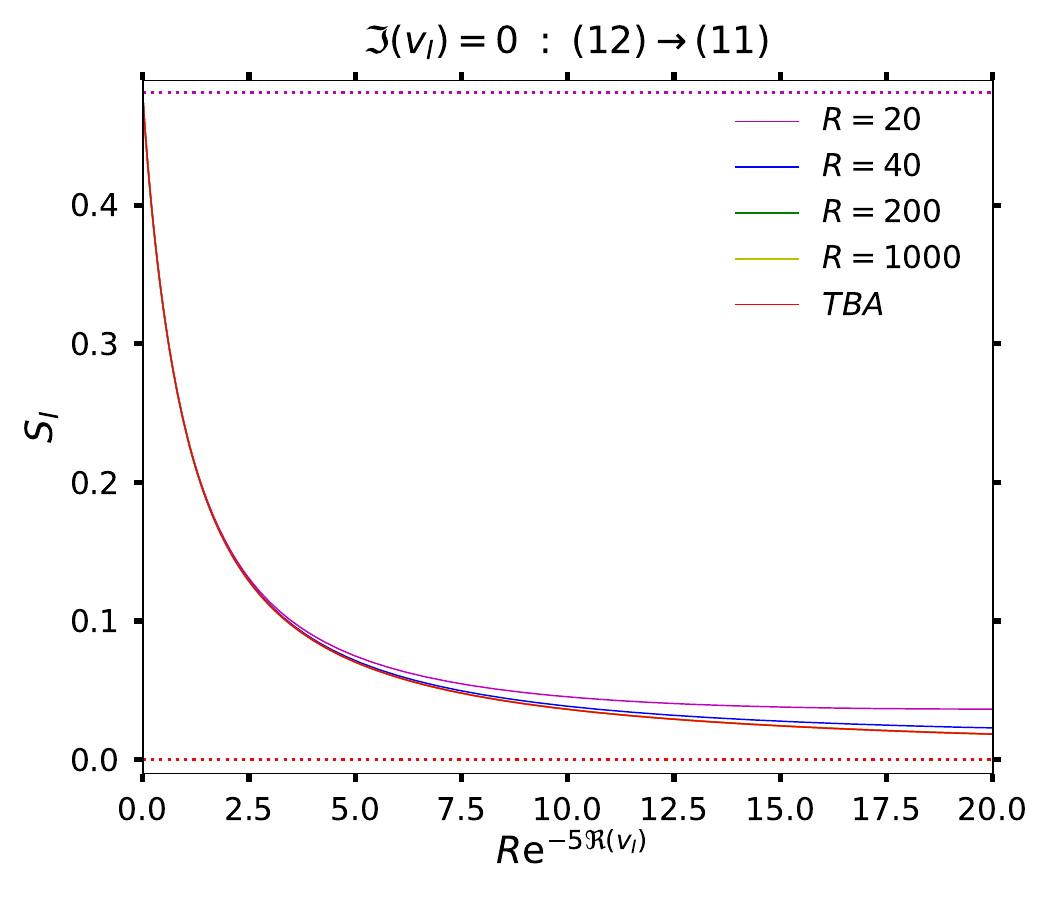}
\end{center}
\end{minipage}%
\begin{minipage}{0.5\linewidth}
\begin{center}
\includegraphics[width=\linewidth]{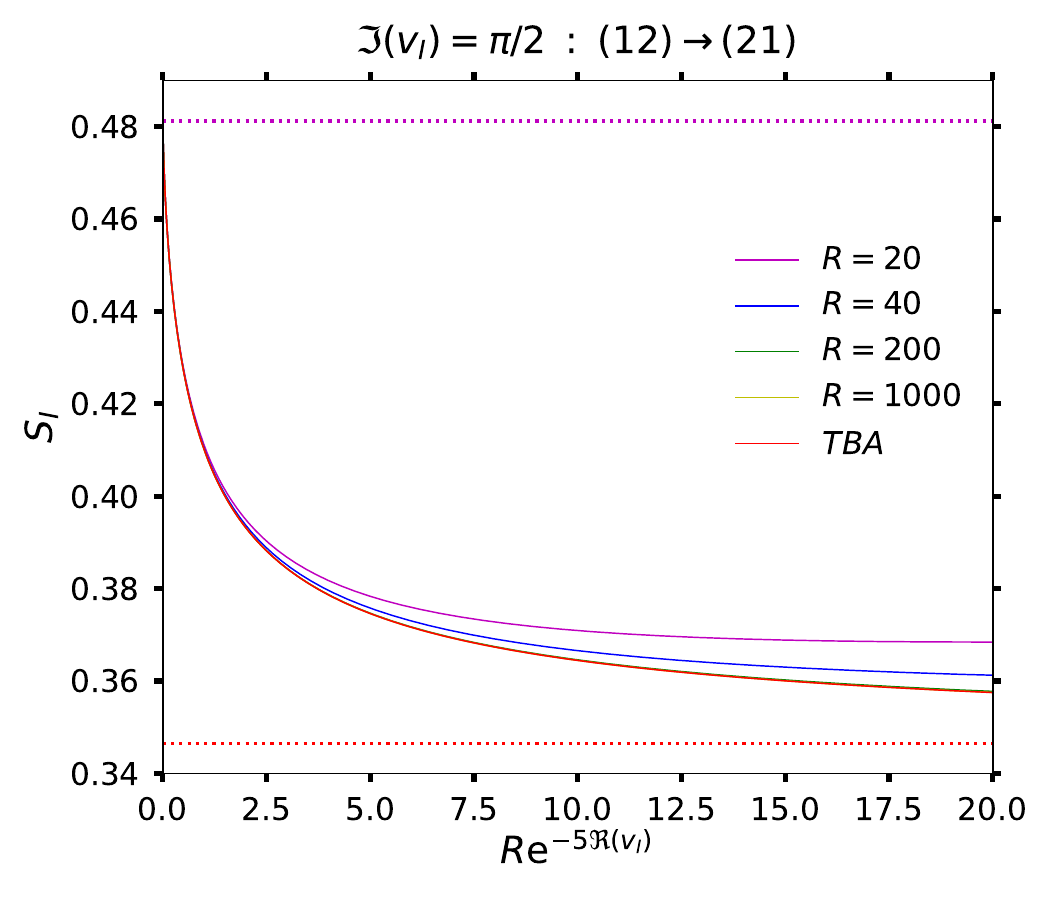}
\end{center}
\end{minipage}
\caption{Comparison between finite-size NLIE and  TBA results (expected to coincide with the  infinite size NLIE results). The left panel - describes the RG flow from  (12) to (11), and the  right from (12) to (21).  }
\label{convergence}
\end{figure}

As mentioned earlier, the NLIE allows us to study finite sizes as well. We illustrate on Figure \ref{convergence} the convergence of the results as $\R$ increases. This convergence is quick enough that, were the system not integrable, direct ab-initio calculations of $\omega^{(J)}$ in the cross-channel would probably be possible.

Starting with the tricritical Ising model we encounter, for many defects,  a phenomenon of level crossings as we vary the crossover scale. This can be explained as follows. The spectrum of the theory without defect is mande of diagonal fields with weights $(\Delta_{11},\Delta_{11}),(\Delta_{22},\Delta_{22})\ldots$. Consider now the case where we have a defect (12). The new spectrum is obtained by fusion, and comprises $(\Delta_{11},\Delta_{12}),(\Delta_{22},\Delta_{21}),(\Delta_{22},\Delta_{23}),\ldots$. 
For tricritical Ising, the corresponding numerical values of the dimensions are 
\begin{eqnarray}\label{levels}
\Delta_{11}+\Delta_{12}={1\over 10},\nonumber\\
\Delta_{22}+\Delta_{21}={19\over 40},\nonumber\\
\Delta_{22}+\Delta_{23}={3\over 40}.
\end{eqnarray}
Hence with defect (12), the lowest dimension is the third value, equal to ${3\over 40}$. If we now consider the flow from (12) to (11), the lowest dimension in the IR will come from $\Delta_{11}+\Delta_{11}=0$:  this is the evolution from the first level in (\ref{levels}) in the UV, not the third one. Hence we expect a level crossing in the data for the ground state of the system as a function of the scaling variable. This crossing can easily be observed in finite systems using e.g. direct diagonalization: it would be interesting to see whether it can be observed within the framework of perturbation theory or the Truncated Conformal Space Approach \cite{Graham:2003nc}. While in many cases a level crossing for the ground-state would correspond to a phase transition, we emphasize here that the model remains gapless and critical throughout the defect RG flow:  the degeneracy at the crossing point can probably be interpreted as the existence of a local zero mode, whose physical significance remains to be elucidated further.

In practice, since the ab-initio curves focus on the ground-state, some  irregularities could be expected as the result of the crossing, but they are barely visible on our data -see   Fig. \ref{fig:TCI_GS}.  Of course, the problem could be avoided by following directly the level corresponding to $\Delta_{11}+\Delta_{12}$: in that case a smooth curve would be obtained indeed. 

\begin{figure}
    \centering
    \includegraphics[width=1.1\linewidth]{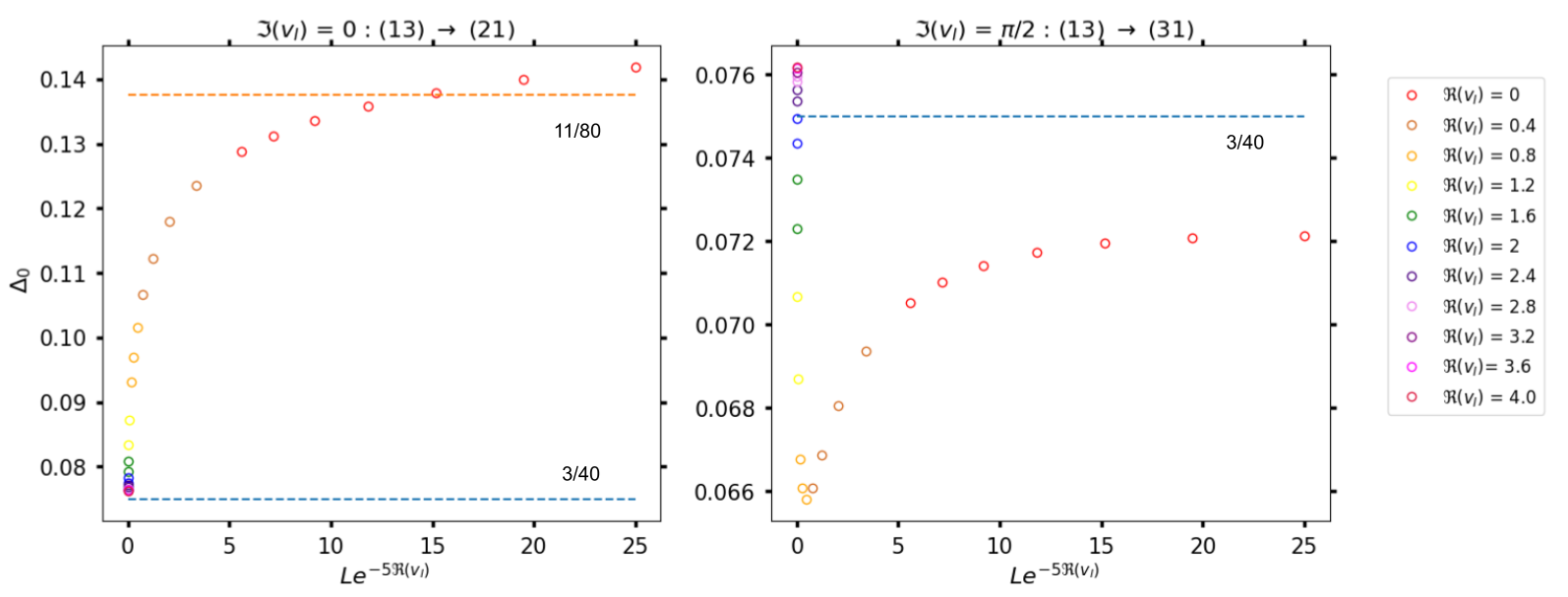}
 \caption{RG flow of the scaling dimension of the ground state for  the tricritical Ising model with defect $J=2$. The ground state energy for the defect Hamiltonian~[Eq.~\eqref{eq:two-imp-ham}] was obtained using Exact Diagonalization   for~$L = 22$ to 28 in steps of 2.~$\Re(v_I)$ was chosen to vary between 0 and 4.0. The left and right panels respectively show the results for the flow from the (13) to (21) and (13) to (31) TDLs   The scaling dimension of the ground state~$\Delta_0$ is a function of~$L, v_I$. The data collapses to a single curve when plotted against the dimensionless variable:~$Le^{-5\Re(v_I)}$. The expected conformal dimensions at the (13), (21) and (31) fixed points are $3/40$, $11/80$ and $3/40$ respectively.  }
    \label{fig:two-imp-tci-flow}
\end{figure}

Finally, we give for completeness in Figure \ref{fig:two-imp-tci-flow} the RG flow of the scaling dimension of the ground state for the tricritical Ising model this time in the flows $(13)\to (21)$ and  $(13)\to (31)$. Note that, in view of the rather complex  form of the Hamiltonian (involving the Jones-Wenzl projectors) we have only used Exact Diagonalization, and thus been limited to rather small sizes. As a result, the  quality of the convergence is rather poor, especially for the right panel. We have in fact  observed  a systematic decrease of the quality of this kind of  data as $p$ increases and corrections to scaling are getting stronger.

\subsection{p = 5: Tetracritical Ising model}

\begin{figure}[H]
\centering
\includegraphics[width = 0.6\textwidth]{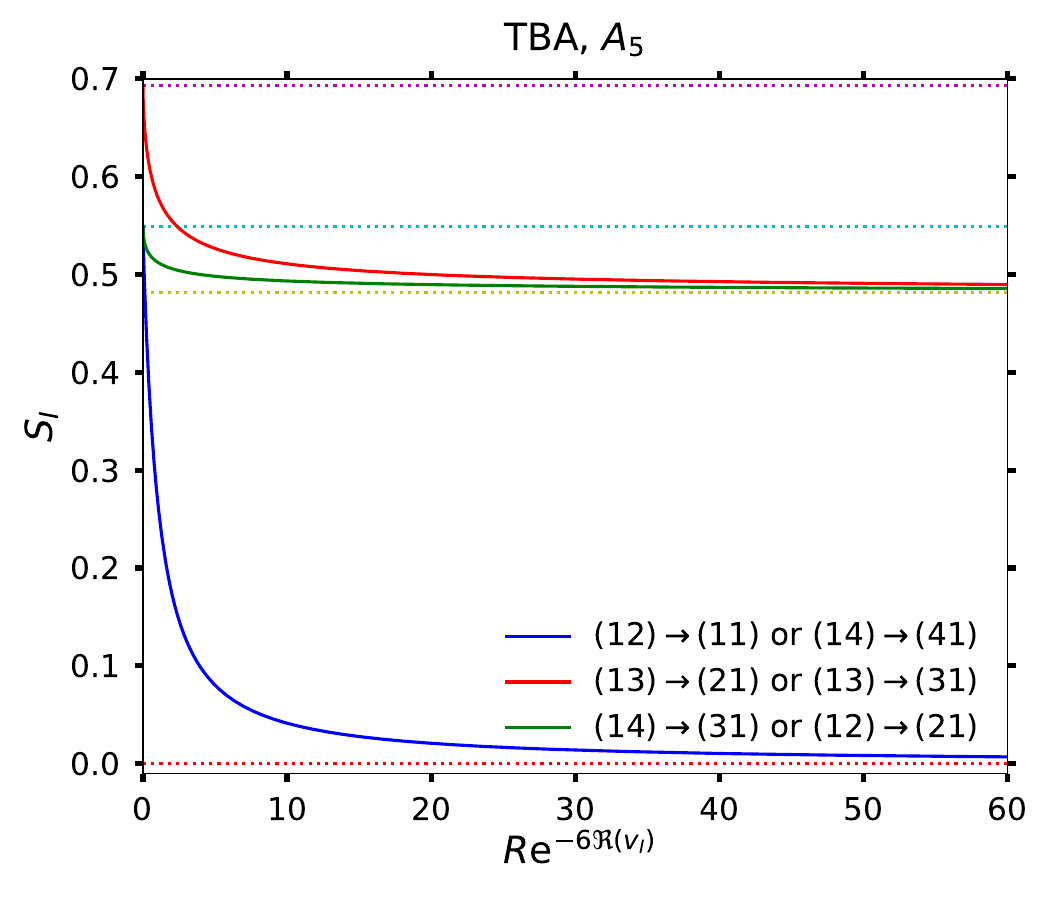}
\caption{Change in the impurity entropy along the RG flow. 
 } 
\label{fig:TBA_A5_flow} 
\end{figure}

It is clear that ab-initio calculations get increasingly  complicated as the value of $p$ is increased: the size of the Hilbert space gets larger, while the number of levels associated with small values of the exponents $h_{(rs)}$ means there can be several crossings as one evolves from one defect to another. On the other hand, the TBA and NLIE approaches work equally well, and there is no reason to doubt that there would be significant difference with good quality ab-initio calculations. Just to illustrate the type of results one can obtain, we provide  in Figure \ref{fig:TBA_A5_flow} several examples for the case $p=5$, which is believed to describe the tetracritical Ising model. 

\subsection{The large $p$ limit and the Kondo problem}

It will not have escaped the knowledgeable reader that the TBA describing the flow from $(1,1+J)$ to $(J,1)$ bears a lot of resemblance to the TBA for the anisotropic Kondo model with an impurity of spin $J$ \cite{Fendley_1996}. The latter would have the exact same expression for the impurity free-energy (\ref{basicint}) but a different TBA diagram - $D_{p+1}$ instead of $A_{p-2}$, with three additional  nodes forming a ``fork'' at the massless end. In this case, the calculation of the impurity free-energy would lead to 
\begin{equation}
\text{f}_I^{(J)}=-T\ln (1+J),
\end{equation}
corresponding to the fact that a defect of spin $J$ at infinite rapidity $v$ behaves like a free additional spin $J$ embedded in the chain. The flow from $(1,1+J)$ to $(J,1)$ in the minimal models  can then be interpreted as the usual Kondo flow from spin $J$ to spin $J-1$, with partial screening due to the absorption of one spinon by the impurity spin \cite{Hewson_1993}. Scattering matrices describing this process are of course well known - see e.g. \cite{PhysRevLett.71.2485, Fendley_1996}. The scattering matrices for the flow from $(1,1+J)$ to $(J,1)$ in the case of minimal models follow from the usual quantum group restriction \cite{cmp/1104200706}. 

As discussed e.g. in \cite{Reshetikhin:1993wm} or \cite{Fendley1993,Fendley1994}, the TBA for the  sine-Gordon model related, via truncation of the diagram, to the TBA for the massive minimal model, corresponds to  a bare coupling which is equal to the square root of the coupling constant in the un-truncated case. This is why, for instance, a minimal model perturbed by the (bulk) conformal field $\Phi_{(13)}$ and a negative coupling constant (so that the flow is massless) is related with the sine-Gordon model perturbed by the usual cosine term but with a {\sl purely imaginary amplitude}. The same holds for defect  flows in minimal models and their relationship with flows in the imaginary anisotropic Kondo model - that is, the Kondo model with a purely imaginary coupling constant. Hence the flow from $(1,1+J)$ to $(1+J,1)$ is nothing but the RSOS version of the flow between the UV fixed point of spin $J$ Kondo (i.e. a decoupled spin $J$) and whatever IR fixed point this model flows to when the coupling is purely imaginary. We plan to discuss this point in greater detail elsewhere.

\bigskip
To get back to the RSOS models now, in the limit $p\to\infty$, the effect of the truncation should be negligible, and properties of the RSOS and Kondo model should become identical. Indeed, for the ordinary flow from $(1,1+J)$ to $(J,1)$ we do indeed see that the ratio $g_{(1,1+J)}/g_{(J,1)}\to {1+J\over J}$ as $p\to\infty$ corresponding to the usual Kondo screening (see \cite{Gaiotto_2021} for related remaks). We also observe  that the ratio $g_{(1,1+J)}/g_{(1+J,1)}\to 1$ as $p\to\infty$, which suggests that {\sl there is no screening in the imaginary  Kondo problem} and that the spin $J$ is decoupled in the IR as it is in the UV. In fact, because the flow of entropies is monotonic in the RSOS models, this strongly suggests  in particular that there is no Kondo effect in the scaling limit for purely imaginary coupling and an isotropic interaction \cite{Nakagawa_2018,Kattel2024}. This point, too, will be elaborated elsewhere.

\section{Flows on TDLs and Fusion} 
\label{sec:fusion}

A good part of the interest of  TDLs in a CFT is that they can be manipulated in various ways. In particular they can be fused, producing fusion categories that are the key to our understanding of non-invertible symmetries. An interesting question is, what happens to this fusion when we implement an RG flow on the defect lines. It is not clear how much to expect in general: as soon as the perturbation is turned on, the lines cease to be topological, and new effects are expected to appear where the shape of the lines might start playing a role. Nonetheless, within the integrable formalism, it is perfectly  possible to introduce two defect lines as two lines with equal (or different) impurity  spectral parameters and spin $J$, and explore the resulting physics. Note that in this set-up the distance between the lines  will not play a role since they can be moved around using the Yang-Baxter equation: this effect is not an artefact of integrability, but matches a similar property in the continuum limit since the perturbations are purely chiral.

Putting two defect lines of type $(1s)$ on the lattice is well known to reproduce the fusion expected from the CFT
\cite{Sinha:2023hum,Sinha2024}. This is in fact an obvious consequence of the algebraic construction in \cite{Belletete2023} and the fact that the lines are topological on the lattice as well. In general - and still remaining at fixed points -  it is not totally clear that fusion and continuum limits commute: if we have two defect lines with $v_I={i\pi\over 2}$, each reproducing on its own  ${\cal D}_{(21)}$ in the continuum limit, does the two defect system flow to ${\cal D}_{(21)}\times {\cal D}_{(21)}={\cal D}_{(11)}+{\cal D}_{(31)}$ in the continuum limit? This question was partly explored in \cite{Sinha:2023hum} in the case of the three-state Potts model  and will be discussed in more detail in \cite{Sinha2024} - but we would like to consider it here briefly as well from a more formal point of view.

It turns out indeed that the integrable formalism provides us with a tool to gain some further understanding of fusion. This relies on the fact that    the transfer matrices $T^{(J)}$ introduced earlier in this paper satisfy a set of 
 bilinear relations
\begin{equation}
T^{(k)}_{\left[\frac{d}{2}+\frac{\ell}{4}\right]} T^{(k+\ell)}_{\left[-\frac{d}{2}-\frac{\ell}{4}\right]} = T^{(k-d)}_{\left[\frac{\ell}{4}\right]} T^{(k+d+\ell)}_{\left[-\frac{\ell}{4}\right]}+T^{(d-1)}_{\left[\frac{k+1}{2}+\frac{\ell}{4}\right]} T^{(d-1+\ell)}_{\left[-\frac{k+1}{2}-\frac{\ell}{4}\right]},
\label{Bilinearev0}
\end{equation}
where we have introduced  the notation $T^{(k)}_{\left[q\right]}= T^{(k)}(v_I+\im q\gamma)$ (recall $\gamma={\pi\over p+1}$), $k,\ell,d$ are positive integers with the restriction that $1 \leq d \leq k< p$. The super-indices denote symmetrically fused representations, of which $k=0$ and $k=1$ are the trivial and the fundamental ones, respectively, and are given by $T^{(0)}$ and $T^{(1)}$ discussed earlier.

Note that  a subset of equation (\ref{Bilinearev0}) with $\ell=0$, $d=1$ and $k=J$ provide the original $T$-system familiar in the literature :
\begin{equation}
T^{(J)}_{\left[\frac{1}{2}\right]} T^{(J)}_{\left[-\frac{1}{2}\right]} = T^{(J-1)}_{\left[0\right]} T^{(J+1)}_{\left[0\right]}+T^{(0)}_{\left[\frac{J+1}{2}\right]} T^{(0)}_{\left[-\frac{J+1}{2}\right]}, \label{tsystem}
\end{equation}
which in turn can be used to define the $y$-system:
\begin{equation}
Y^{(J)}(v_I)=\frac{T^{(J)}_{\left[\frac{1}{2}\right]} T^{(J)}_{\left[-\frac{1}{2}\right]}}{T^{(0)}_{\left[\frac{J+1}{2}\right]} T^{(0)}_{\left[-\frac{J+1}{2}\right]}} = \frac{T^{(J-1)}_{\left[0\right]} T^{(J+1)}_{\left[0\right]}}{T^{(0)}_{\left[\frac{J+1}{2}\right]} T^{(0)}_{\left[-\frac{J+1}{2}\right]}}+1=y^{(J)}(v_I)+1, \label{ysystem}
\end{equation}
for $J=1,~2,~\ldots,p-2$.
The first equality in (\ref{ysystem}) allows us to extract the eigenvalue $\Lambda^{(J)}(v_I)$ along the line $\Im v_I =0$ in terms of the $Y^{(J)}(v_I)$. The last two equalities in (\ref{ysystem}) re-phrases the T-system (see eq. (\ref{ty-system}) and allows to establish non trivial relations among $y,~Y$'s in the form of the NLIE's (\ref{NLIE}).

\begin{figure}[htb]
\begin{minipage}{0.5\linewidth}
\begin{center}
\includegraphics[width=0.8 \linewidth]{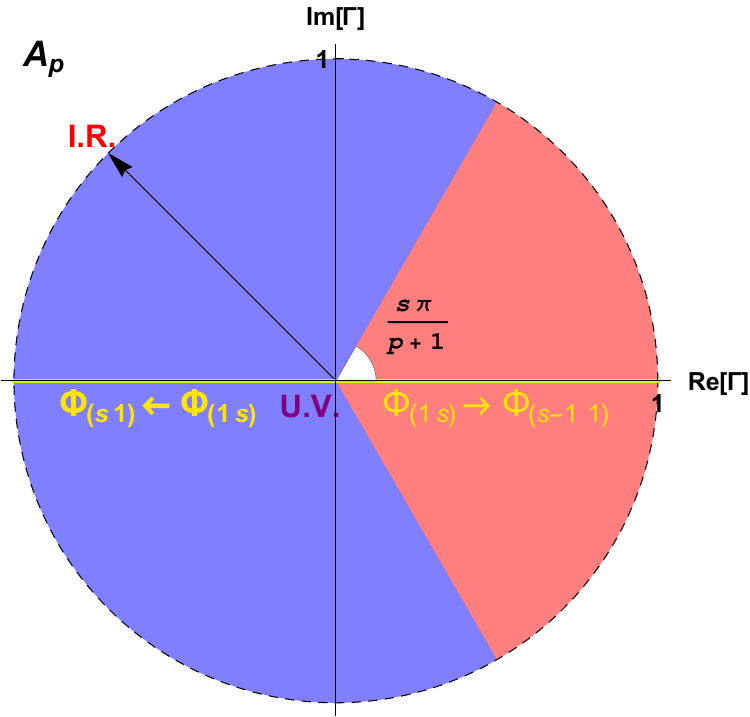}\\
\includegraphics[width=0.8\linewidth]{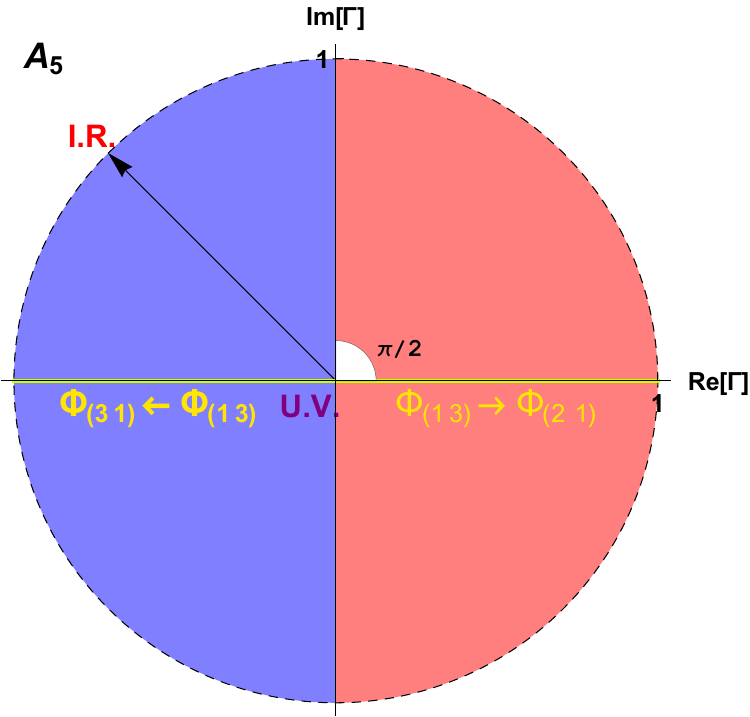}
\end{center}
\end{minipage}%
\begin{minipage}{0.5\linewidth}
\begin{center}
\includegraphics[width=0.8\linewidth]{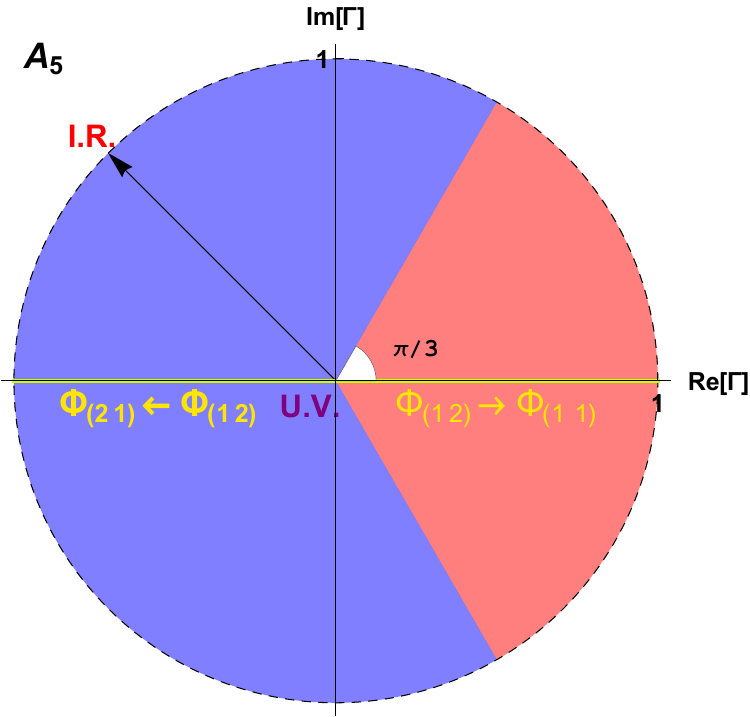}\\
\includegraphics[width=0.8\linewidth]{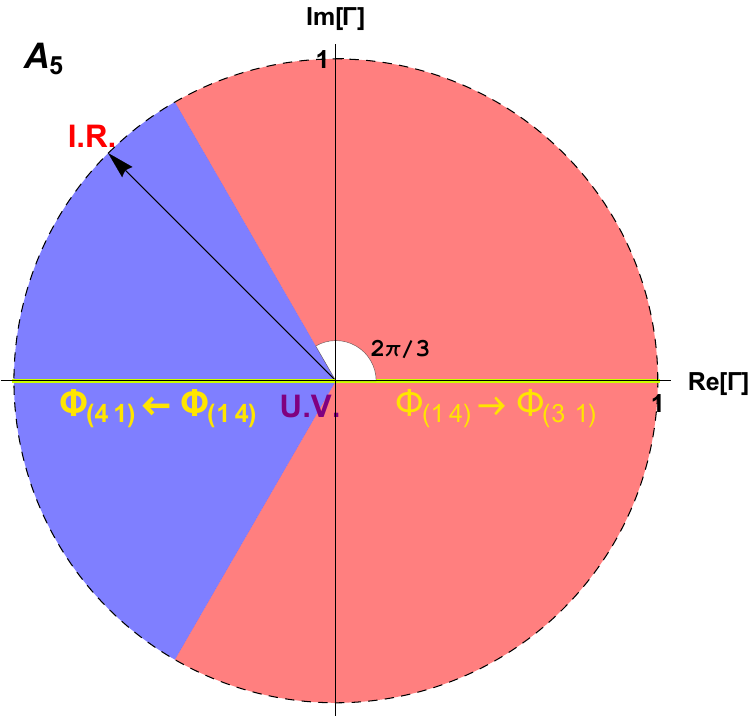}
\end{center}
\end{minipage}
\caption{a) Diagrammatic illustration  of (\ref{thumbs}) in the $\Gamma$-plane (see eq. (\ref{eq:H_pert}). The U.V. fixed point sits  the origin and can flow to two different I.R. fixed points (schematically situated on  the perimeter of the circle)  identified by  the shading. The two different regions always include an hermitian incarnation of (\ref{eq:H_pert})  on the real line, where the sign of the perturbation determines the fate of the flow. b) Tetra-critical Ising with $J=2$. c) Tetra-critical Ising with $J=3$. d) Tetra-critical Ising with $J=4$.} 
\label{Thumbs}
\end{figure}

Now, the key idea is to use  eqs. (\ref{Bilinearev0}) to study  fusion of the TDLs. An immediate obstacle to doing so is the fact  that the arguments of the transfer matrices on the left are shifted by imaginary quantities in the complex plane. This doesn't matter for the $(1s)$ fixed points since for these the spectral parameter $v_I$ is sent to infinity anyhow. For for the $(r1)$ fixed points as well as the RG flows, this creates  a  potential problem. However, while the Hamiltonians in the direct channel (or the transfer matrices in the cross-channel) are not hermitian away from the lines $\Im(v_I)=0$ or $\Im(v_I)=\pm {\pi\over 2}$, it turns out nevertheless that the flows are quite stable within entire domains of the complex $v_I$ plane, with the result that 
\begin{equation}
(1,s),~s\geq 2 \rightarrow 
\begin{cases}
(s-1,1), &\text{for } |\Im{(v_I)}|< s \pi/(2p+2),\\
(s,1),&\text{for } \frac{\pi}{2}\geq|\Im{(v_I)}|> s \pi/(2p+2), \label{thumbs}
\end{cases}
\end{equation}
(where we recall $\gamma={\pi\over p+1}$.) This is illustrated in  Fig. \ref{Thumbs}. 

In simple terms, we see that the IR fixed point of the flows are stable provided one doesn't stray too far from the lines $\Im(v_I)=0,\Im(v_I)=\pm {\pi\over 2}$: in particular, the shifts in the fusion equations (\ref{tsystem}) can mostly be ``forgotten when'' identifying the corresponding flows.

Now, a crucial point is that  relations (\ref{NLIE}-\ref{omega}) are valid for $v$ inside the analytical non-zero strip(ANZ) $|\Im (v_I)|<\gamma/2$. One can extend this region for the eigenvalue expression, but this again will turn out to be limited. For the time being, whether or not we extrapolate the original strip is irrelevant. It matters, however, that the strips for $ \Lambda^{(J)}(v_I)$ containing $\Im (v_I) =0$ and $\Im (v_I) = \pi/2$ are separated by dense line of zeros of the eigenvalue - which is of course at the root of the different fates of the RG flow.

For $\Im(v_I)=0$, we have the simple result for eigenvalues
\begin{equation}
\Lambda^{(1)}_{\left[\frac{1}{2}\right]} \Lambda^{(1)}_{\left[-\frac{1}{2}\right]} = \Lambda^{(0)}_{\left[0\right]} \Lambda^{(2)}_{\left[0\right]}+\Lambda^{(0)}_{\left[1\right]} \Lambda^{(0)}_{\left[-1\right]}. \label{12times12}
\end{equation}
and the expressions (\ref{NLIE}-\ref{omega}) hold for all terms on the left and right hand-sides.

For $\Im(v_I)=\pm{\pi\over 2}$, the quickest way to proceed  is to make use of the reflection relation. The latter can be obtained from (\ref{Bilinearev0}) by taking  $k+d+\ell = p$ and $d=1$ (so  $\ell = p-1 - k$) and recalling that $T^{(p)}(v)= T^{(-1)}(v)=0$.  We have then
\begin{equation}
T^{(k)}_{\left[0\right]} = T^{(0)}_{\left[k/2\right]}  {\left(T^{(p-1)}\right)}^{-1}_{\left[k/2-(p+1)/2\right]} ~T^{(p-1-k)}_{\left[-(p+1)/2\right]} = \Pi ~T^{(p-1-k)}_{\left[ -(p+1)/2\right]}, \label{reflection}
\end{equation}
where $\Pi = {(-1)}^{ \frac{p \R}{2}}{\cal R}$ with ${\cal R}$ the height reflection. It follows that we can use\footnote{We note that eigenvalues in fact  occur in pairs (for $k$ even) or in pairs of  opposite sign and same modulus (for $k$ odd), so has to be careful with interepreting this  identities. The problem however disappears when considering products of eigenvalues as we do below.}  
$\Lambda^{(k)}(v_I)=\Lambda^{(p-1-k)}\left(v_I-{i\pi\over 2}\right)$ in the fusion equations  so, setting $v_I=\Re(v_I)+{i\pi\over 2}$, we obtain  the following system
\begin{equation}
\Lambda^{(p-2)}_{\left[\frac{1}{2}\right]} \Lambda^{(p-2)}_{\left[-\frac{1}{2}\right]} = \Lambda^{(p-3)}_{\left[0\right]} \Lambda^{(p-1)}_{\left[0\right]}+\Lambda^{(0)}_{\left[\frac{p-1}{2}\right]} \Lambda^{(0)}_{\left[-\frac{p-1}{2}\right]}. \label{21times21}
\end{equation}
where the argument of the eigenvalues on both sides must be taken as $\Re(v_I)$ instead of $v_I$. Expressions (\ref{NLIE}-\ref{omega}) with $J=p-2$ then can be reliably used. 

At first sight, a relation like eq. (\ref{21times21}) looks like an extension of the fusion ${\cal D}_{(12)}\times {\cal D}_{(12)}={\cal D}_{(11)}+{\cal D}_{(13)}$ to the whole RG flow. One has to be careful however with the matter of bulk terms $u^{(J)}$ briefly mentioned already in section \ref{sec:NLIE} (see in particular eq. (\ref{bulk})): relation (\ref{21times21}) will translate into a similar identity for the interesting and universal factors $\omega^{(J)}$ if and only if  the bulk term of the l.h.s. is equal to both bulk terms on the r.h.s. Surprisingly, this is true in some cases but not in others.

Specifically, the following identity holds for all $J\geq0$:
\begin{equation}\label{goodId}
u^{(J)}(v+\im \gamma/2)+u^{(J)}(v-\im \gamma/2) = \log \sinh(v+\im (J+1) \gamma/2) \sinh(v-\im (J+1) \gamma/2).
\end{equation}
while, for $J>1$, we also have 
\begin{equation}
u^{(J-1)}(v)+u^{(J+1)}(v) = \log \sinh(v+\im (J+1) \gamma/2) \sinh(v-\im (J+1) \gamma/2).
\end{equation}
Therefore, for $J>1$ all bilinear terms in the T-system (\ref{tsystem}) have the same extensive behavior. Taking  $J=p-2$ and $\Re(v_I)=0$, we find that (\ref{21times21}) translates into $g_{21}g_{21}=g_{11}+g_{31}$, a necessary result for the proper identification of ${\cal D}_{(21)}$ \cite{Sinha2024}. More generally, an identity of this type holds for the $\omega$ factors and thus the RG dependent entropies. Note this implies that fusion and RG commute: if we first fuse the defects in the UV we get $(12)\times (12)=(11)+(13)$. Following the RG in the region where $(12)$ flows to $(21)$ we obtain, on the l.h.s., $(21)\times (21)$. On the r.h.s. we get on the other hand $(11)\to (11)$ and $(13)\to (31)$, which add up to $(21)\times (21)$ indeed. 

For $J=1$ however, something different happens because the correct expression for $u^{(0)}(v)$ is not obtained by extending (\ref{bulk}) to $J=0$: rather, we have $u^{(0)}(v_I)=\ln\sinh(v_I)$. It follows that, in the scaling limit and for  finite $v_I$   
\begin{equation}
\frac{\Lambda^{(0)}(v_I) \Lambda^{(2)}(v_I)}{\Lambda^{(1)}(v_I+\im \gamma/2) \Lambda^{(1)}(v_I-\im \gamma/2)} \to 0,~\text{when}~\R \to \infty, 
\end{equation}
Hence,  at the IR fixed  point, the contribution of the first term on  the rhs of (\ref{12times12}) vanishes, while the bulk term of the l.h.s. is still equal to the bulk term of the second part of the r.h.s. thanks to (\ref{goodId}). This, in fact, can again be interpreted as commutation of fusion and RG. Indeed, in the UV we have $(12)\times (12)=(11)+(13)$. In this region of couplings, we have $(12)\to (11)$, so under RG the l.h.s. flows to $(11)\times (11)=(11)$. Naively, one would then encounter a paradox with the r.h.s. since $(11)\to (11)$ and $(13)\to (21)$. The point however is that the second term is in fact killed by the bulk contribution: $(11)\times (13)\to (01)\times (21)$, so the $(13)$ channel eventually disappears.

\section{Summary and Outlook}
\label{sec:concl}
To summarize, in this work, we analyzed RG flows ``on'' topological defect lines (more precisely, starting and ending on TDLs)  in unitary minimal models of conformal field theories. We presented a family of parameter-dependent integrable lattice models\footnote{Note that, while we relied mostly on the RSOS version, the translation to the anyonic-chain version is straightforward \cite{belletete2020topological}.} making possible Bethe-ansatz and ab-initio numerical computations of the flows between the different fixed points. While of interest in itself, this study is also an essential component of the program of identifying TDLs in lattice models, since there is now overwhelming evidence that defects such as $(r1)$ can only be obtained as end-points of RG flows of the type studied here.

While  flows on single lines are now well understood, the combination of fusion and flows needs more work, both from a lattice and a field theoretic point of view. Section 
\ref{sec:fusion} provided a promising avenue for further investigation within the integrable formalism. More ab-initio calculations would also be useful, in particular those involving two defect lines where the spectral parameters are not shifted and remain physical. This will be discussed elsewhere \cite{Sinha2024}. Similarly, a deeper understanding of the bulk terms  $u^{(J)}$ from a perturbative point of view might shed some light on the fate of different terms on the r.h..s of fusion equations such as (\ref{12times12}).

Finally, while we restricted our attention to RG flows between defect lines in the same theory, the same framework can be  generalized to investigate the  fate of TDLs when there is a {\sl bulk} flow, be it  between different CFTs, or between a CFT and a massive phase \cite{Chang:2018iay}. We hope to get back to this question soon.

\section*{Acknowledgements}
The authors thank P. Fendley and S. Lukyanov for discussions and  J. Bellet\^ete, A. Gainutdinov, J.L. Jacobsen and F. Yan  for related collaborations. The work of H.S. was supported by the French Agence Nationale de la Recherche (ANR) under grant ANR-21- CE40-0003 (project CONFICA). A.R. was supported by a grant from the Simons Foundation (825876, TDN).

\newpage

\bibliography{library_1}

\end{document}